\documentclass{article}

\usepackage{arxiv}
\usepackage[utf8]{inputenc} % allow utf-8 input
\usepackage[T1]{fontenc}    % use 8-bit T1 fonts
\usepackage{hyperref}       % hyperlinks
\usepackage{url}            % simple URL typesetting
\usepackage{booktabs}       % professional-quality tables
\usepackage{amsfonts}       % blackboard math symbols
\usepackage{nicefrac}       % compact symbols for 1/2, etc.
\usepackage{microtype}      % microtypography
\usepackage{booktabs} % For formal tables
\usepackage{bbold}
\usepackage{amsmath}
\usepackage{pdfpages}
\usepackage{subfigure}
\usepackage{natbib}
\usepackage{bbold}
\usepackage{amsmath}
\usepackage{pdfpages}
\usepackage{url}
\usepackage{subfigure}
\usepackage{enumitem}
\usepackage{multirow}
\usepackage{colortbl}

\usepackage[ruled]{algorithm2e} % For algorithms

\SetAlFnt{\small}
\SetAlCapFnt{\small}
\SetAlCapNameFnt{\small}
\SetAlCapHSkip{0pt}
\IncMargin{-\parindent}
\newcommand{\todo}[1]{\textcolor{red}{#1}}

\title{Semantic Table Retrieval using Keyword and Table Queries}  

\author{
  Shuo Zhang\\
  University of Stavanger \\
  \texttt{shuo.zhang@uis.no} \\
   \And
 Krisztian Balog \\
  University of Stavanger\\
  \texttt{krisztian.balog@uis.no} \\
}

\begin{document}
\maketitle

\begin{abstract}
Tables on the Web contain a vast amount of knowledge in a structured form. To tap into this valuable resource, we address the problem of table retrieval: answering an information need with a ranked list of tables. 
We investigate this problem in two different variants, based on how the information need is expressed: as a keyword query or as an existing table (``query-by-table'').
The main novel contribution of this work is a semantic table retrieval framework for matching information needs (keyword or table queries) against tables.  Specifically, we (i) represent queries and tables in multiple semantic spaces (both discrete sparse and continuous dense vector representations) and (ii) introduce various similarity measures for matching those semantic representations.  We consider all possible combinations of semantic representations and similarity measures and use these as features in a supervised learning model.
Using two purpose-built test collections based on Wikipedia tables, we demonstrate significant and substantial improvements over state-of-the-art baselines.
\end{abstract}


%\thanks{\new{This is a revised and extended version of \citet{Zhang:2018:AHT}. }\newline
%  Authors' addresses: Shuo Zhang, Bloomberg, 3 Queen Victoria St, London EC4N 4TQ, Unite Kindom.
%  
%  Krisztian Balog, Dept. of Electrical Engineering and Computer Science, NO-4036 Stavanger, Norway.}

%\renewcommand{\shortauthors}{S. Zhang et al.}

\section{Introduction}

Tables are a powerful, versatile, and easy-to-use tool for organizing and working with data.  Because of this, a massive number of tables can be found ``out there,'' on the Web or in Wikipedia, representing a vast and rich source of structured information.  Recently, a growing body of work has begun to tap into utilizing the knowledge contained in tables.  A wide and diverse range of tasks have been undertaken, including but not limited to 
(i) searching for tables (in response to a keyword query~\citep{Cafarella:2008:WEP,Cafarella:2009:DIR,Venetis:2011:RST,Pimplikar:2012:ATQ,Balakrishnan:2015:AWP,Nguyen:2015:RSS} or a seed table~\citep{DasSarma:2012:FRT}), 
(ii) extracting knowledge from tables (such as RDF triples~\citep{Munoz:2014:ULD}), and
(iii) augmenting tables (with new columns~\citep{DasSarma:2012:FRT,Cafarella:2009:DIR,Lehmberg:2015:MSJ,Yakout:2012:IEA,Bhagavatula:2013:MEM,Zhang:2017:ESA}, rows~\citep{DasSarma:2012:FRT,Yakout:2012:IEA,Zhang:2017:ESA}, cell values~\citep{Ahmadov:2015:THI}, or links to entities~\citep{Bhagavatula:2015:TEL}). 
Searching for tables is an important problem on its own, in addition to being a core building block in many other table-related tasks.  
Yet, up until recently it has not received due attention, and especially not from an information retrieval perspective. 
Our work, which has been published in~\citep{Zhang:2018:AHT} and is being extended in this paper, was a first attempt at aiming to fill that gap, and has spurred interest in table retrieval ~\citep{Li:2019:TNW, M:2019:ITR, Bagheri:2020:ALM, Shraga:2020:PBR, Shraga:2020:AHT, Chen:2020:TSU, Shraga:2020:WTR,Pyreddy:1997:TAS,Pinto:2002:PDB,Liu:2007:TAT,Wei:2006:TEA}.  In this study, we further extend semantic table retrieval to support table-based search scenarios as well.  Table-based search, despite its practical utility, has not been extensively explored to date.
%\todo{$\Leftarrow$ TODO(SZ): complete list with other papers citing our WWW work.} \todo{TODO(KB): One more sentence stating the importance and contributions of this extension.}

\begin{figure}[t]
   \centering
   \includegraphics[width=0.5\textwidth]{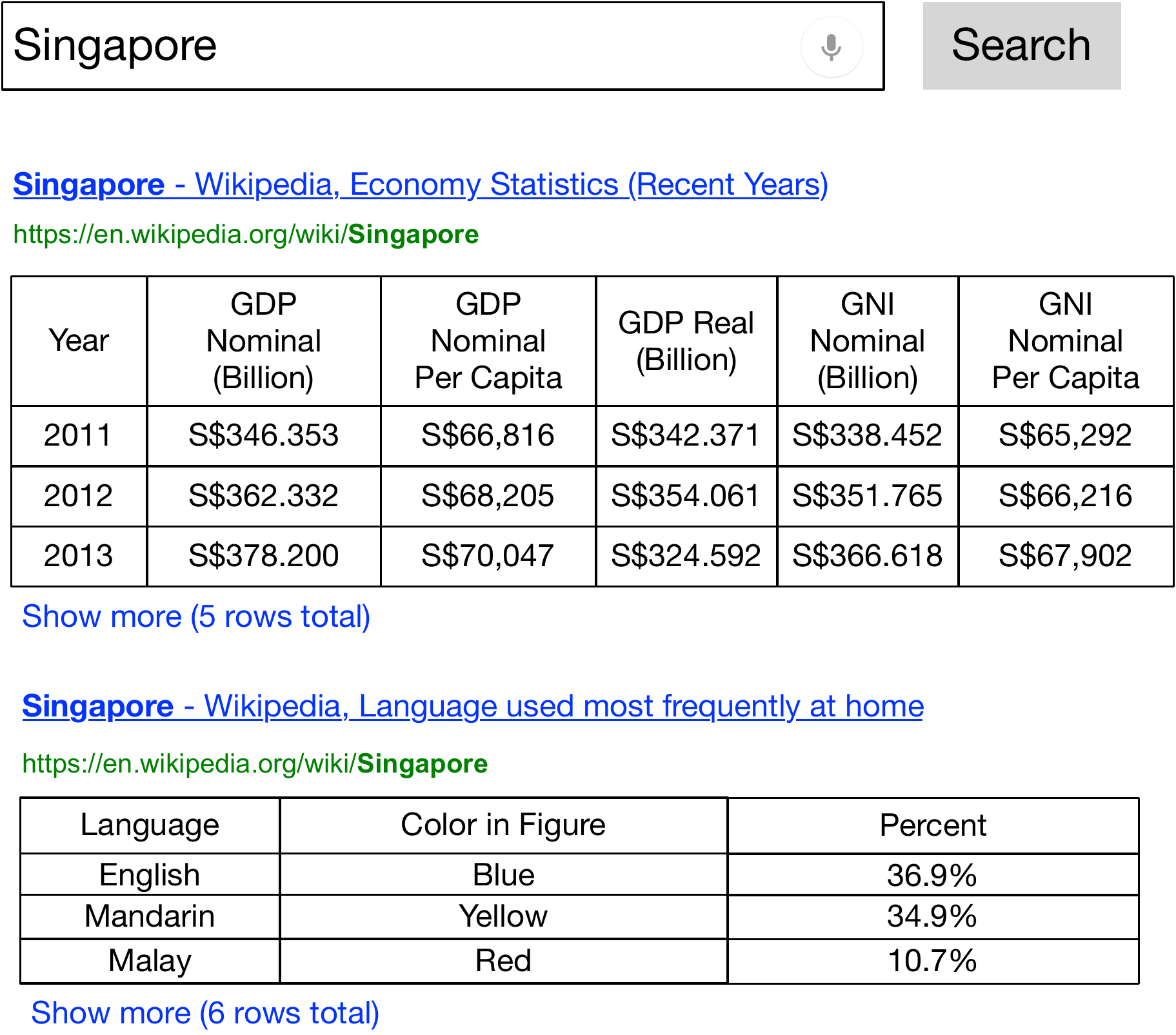} 
   \caption{Keyword-based table search: given a keyword query, the system returns a ranked list of tables.}    \label{fig:tableretrieval1}
\end{figure}

\begin{figure}[t]
   \centering
   \includegraphics[width=0.6\textwidth]{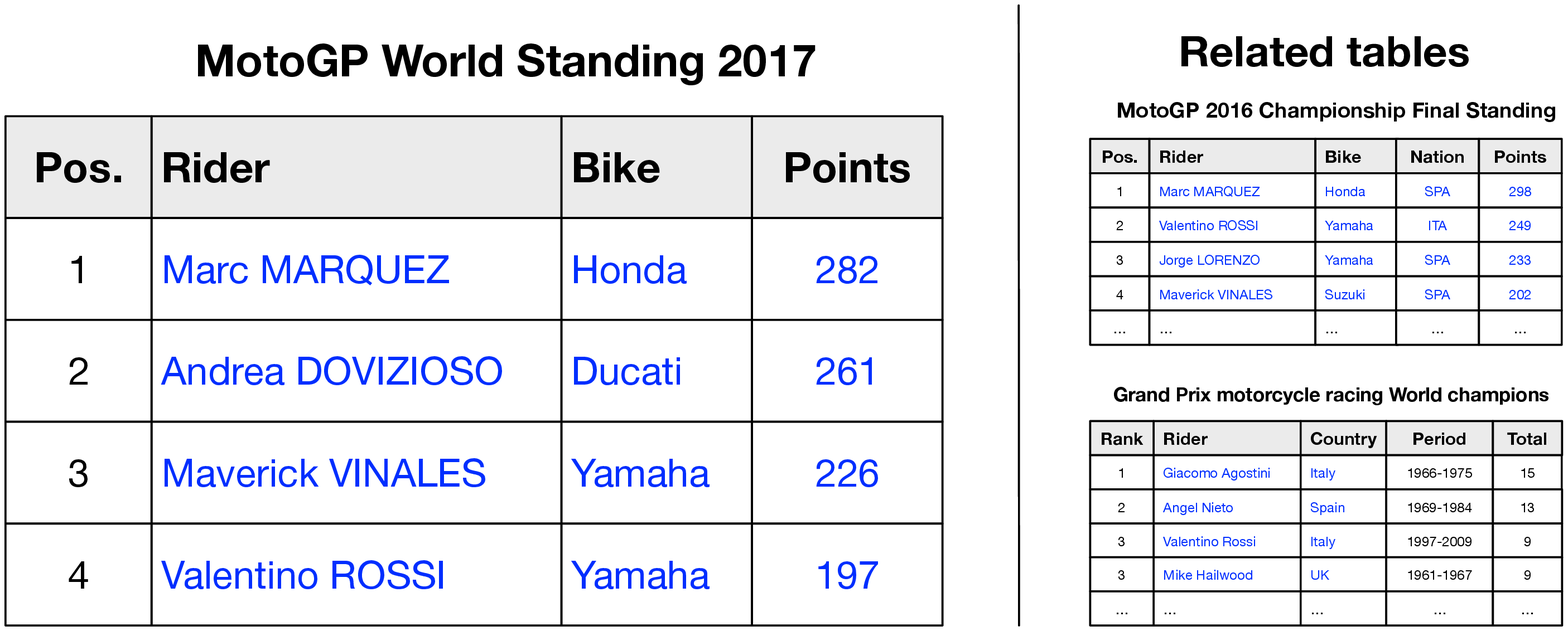} 
   \caption{Query-based table search: given an input table, the system returns a ranked list of tables.}    
   \label{fig:tableretrieval}
\end{figure}

We address the task of \emph{table retrieval}, that is, the problem of generating a ranked list of tables in response to an information need, in two particular flavors:
(i) \emph{keyword-based search}, where the information need is specified as a keyword query, and
(ii) \emph{table-based search}, where an existing table is used as input. 
The former task corresponds to a classical ad hoc search scenario, where tables are sought  for a particular purpose or need.
The latter task resembles more of a recommendation problem, where the user is not required to explicitly formulate a query.  Instead, we suggest tables that contain related information (e.g., additional entities and/or attributes) that could potentially complement the table the user is currently working on.  This ``query-by-table'' paradigm could be helpful, for example, in equipping spreadsheet applications with a smart assistance feature for finding related content.  Alternatively, it could be implemented as a browser plugin that can be activated upon encountering a table on a webpage to find related tables (e.g., for comparison or fact validation).
See Figures~\ref{fig:tableretrieval1} and~\ref{fig:tableretrieval} for an illustration. 
%

%We additionally propose a novel table search paradigm, referred to as \emph{query-by-table}: given an input relational table, return a ranked list of relevant 
%Figure~\ref{fig:tableretrieval} illustrates the idea. 

It should be acknowledged that table retrieval is not an entirely new research problem; in fact, it has been around for a while in the database community (also known there as \emph{relation ranking})~\citep{Cafarella:2008:WEP,Cafarella:2009:DIR,Venetis:2011:RST,Bhagavatula:2013:MEM}.
However, public test collections and proper evaluation methodology are lacking, in addition to the need for better ranking techniques.
Tables can be ranked much like documents, by considering the words contained in them~\citep{Cafarella:2008:WEP, Cafarella:2009:DIR, Pimplikar:2012:ATQ}.  Ranking may be further improved by incorporating additional signals related to table quality.  Intuitively, high quality tables are topically coherent; other indicators may be related to the pages that contain them (e.g., if they are linked by other pages~\citep{Bhagavatula:2013:MEM}).
However, a major limitation of prior approaches is that they only consider lexical matching between the contents of tables and queries. 
This gives rise to our main research objective: 
\emph{Can we move beyond lexical matching and improve table retrieval performance by incorporating semantic matching?}

In this paper, we introduce the \emph{semantic table retrieval (STR)} framework to handle matching in different semantic spaces in a uniform way.  It hinges on the idea of modeling both the table and the input (keyword or table) query as sets of semantic vectors.  Specifically, we consider two main kinds of semantic representations: (i) sparse discrete representations based on entities, and (ii)  continuous vector representations of words and of entities (i.e., word and graph embeddings).  
 We propose two general strategies (early and late fusion), yielding four different measures for computing the similarity between queries and tables based on their semantic representations.  These different similarity scores are then combined in a learning-to-rank framework, together with features aimed capturing general table characteristics. 

As mentioned above, another key area where prior work has insufficiencies is evaluation.  First, there is no publicly available test collection for this task.  Second, evaluation has been performed using set-based metrics (counting the number of relevant tables in the top-$k$ results), which is a very rudimentary way of measuring retrieval effectiveness.
We address this by developing a purpose-built test collections, comprising of 1.6M tables from Wikipedia, and a set of queries with graded relevance judgments.
For both tasks, we develop strong baselines by assembling a rich set of features from prior work and combining them in a learning-to-rank framework.  While all individual features are taken from the literature, the compilation of the feature sets underlying our baselines constitutes an important contribution.  We demonstrate that these strong baselines substantially outperform the best approaches known in the literature. 

Concerning the effectiveness of our STR framework, our findings are as follows.
For keyword-based search, we show that the semantic matching methods we propose can significantly and substantially improve retrieval performance over the strong baseline. 
For table-based search, our proposed approach is on par with the respective strong baseline.  Importantly, this level of performance is reached without requiring the extensive feature engineering that the baseline does.  %Moreover, our analysis reveals that cross-element matching, while seemingly unintuitive, can indeed be beneficial.  
We further demonstrate that retrieval performance increases as the input table grows, either horizontally or vertically, which attests to the capability of our table matching framework to effectively utilize larger inputs.

In summary, this paper makes the following contributions:

\begin{itemize}
	\item We formalize the table retrieval task in two specific flavors: keyword-based search and table-based search (Section~\ref{sec:tableret}). 
	\item We develop strong baselines for both tasks by combining elements from prior studies in feature-based supervised learning approaches (Section~\ref{sec:baselines}).
	\item We present a novel semantic matching framework for table retrieval that can effectively perform matching beyond lexical similarity (Section~\ref{sec:sem}).
	\item We develop standard test collections for both keyword-based and table-based search, which involves gathering relevance assessments (Section~\ref{sec:testcoll}).
	\item We conduct an extensive experimental evaluation and demonstrate the effectiveness of our table retrieval framework (Section~\ref{sec:eval}).  We also carry out a thorough analysis leading into valuable insights (Section~\ref{sec:analysis}).
\end{itemize}
Of these, the table-based search paradigm, methods, and evaluation resources as well as the extensive analysis of both keyword-based and table-based search results are novel contributions on top of the work~\citep{Zhang:2018:AHT} this paper extends.

The resources developed within this paper has been publicly available at \url{https://github.com/iai-group/table-retrieval}.	
%\todo{TODO(SZ): We should add link to a GitHub repo.}

\section{Related Work}

An increasing number of studies are addressing various table-related tasks, including table search, table mining, and table augmentation.  Our work concerns table search, which is considered as a fundamental task both on its own and as a component in other tasks.
Additionally, the line of research on exploiting neural embeddings for IR tasks also bears relevance to this study. 

\subsection{Table Search}

\emph{Table search} answers a query with a ranked list of tables. 
Based on the type of the query, table search can be divided into \emph{keyword-based search}~\citep{Cafarella:2008:WEP,Cafarella:2009:DIR,Venetis:2011:RST,Pimplikar:2012:ATQ,Balakrishnan:2015:AWP,Nguyen:2015:RSS} and \emph{table-based search}~\citep{Lehmberg:2015:MSJ, Ahmadov:2015:THI, DasSarma:2012:FRT, Yakout:2012:IEA, Nguyen:2015:RSS, Limaye:2010:ASW, Zhang:2019:RRT}.

%Early work solves this task for keyword queries. 
The WebTables system by \citet{Cafarella:2008:WEP} pioneered keyword-based table search on top of an existing web search engine. The basic idea is to fetch the top-ranked results returned by a web search engine in response to the query, and then extract the top-$k$ tables from those pages.  Further refinements to the same idea are introduced in~\citep{Cafarella:2009:DIR}.
\citet{Venetis:2011:RST} leverage a database of class labels and relationships extracted from the Web, which are attached to table columns, for recovering table semantics. This information is then used to enhance table search.
Using column keywords, \citet{Pimplikar:2012:ATQ} search tables using term matches in the header, body and context of tables, as signals. An example of a keyword-based table search system interface is provided by Google Web Tables.\footnote{https://research.google.com/tables} The developers of this system summarize their experiences in~\citep{Balakrishnan:2015:AWP}.  Their query is not limited to keywords, it can also be a table.  

Our recent work~\citep{Zhang:2018:AHT}, which this article extends, formally introduces the ad hoc table retrieval task: answering a keyword query with a ranked list of tables.  Semantic matching between queries and tables is proposed as a solution to this problem.
As a follow-up, \citet{Li:2019:TNW} train word and entity embeddings utilizing the Wikipedia table corpus and achieve comparable results.
\citet{M:2019:ITR} put forward a context-aware table search method based on the embeddings for attribute tokens. Different from~\citep{Li:2019:TNW}, \citet{M:2019:ITR} find that differentiated types of contexts such as numerical cell values are useful in constructing word embeddings. 
The system has up to 5\% improvement in
NDCG@5 over LM that uses the same context fields but treats them as the same context. 
The trained model can be used to predict different contexts of every table, which further improves table ranking performance.
\citet{Bagheri:2020:ALM} recognize that some queries constitute tokens that are not well observed in the relevant tables, and propose a latent model to project the token-table co-occurrence matrix into latent factor matrics, which can be used for measuring similarities.
This approach improves over the baselines except for STR~\citep{Zhang:2018:AHT}, which was only significantly outperformed using the Keyword variation.
In a similar setting to~\citep{M:2019:ITR}, i.e., by utilizing table columns and attributes, \citet{Shraga:2020:PBR} propose a projection model for table retrieval using table columns as pseudo-relevance feedback.
Our initial work~\citep{Zhang:2018:AHT} explores the use of both extrinsic similarities, such as entity results for keywords, and intrinsic similarities such as word-level similarities.
Similarly, \citet{Shraga:2020:AHT} make a novel use of intrinsic features (passage-based) and extrinsic features (manifold-based table similarities) for table retrieval.
\citet{Chen:2020:TSU} investigate how to encoding tabular content into BERT for generating embeddings, taking table structure and the input length limit of BERT into consideration.
Taking STR~\citet{Shraga:2020:AHT} as the baseline, the BERT-based methods report on performance improves in the range of 4-7\% in terms of NDCT@20.
To bridge table retrieval and end-to-end embedding learning, \citet{Shraga:2020:WTR} suggest MTR, which utilizes of Gated Multimodal Units (GMUs) to learn a joint-representation of the query and the different table modalities. 
\citet{Shraga:2020:WTR} further extend table retrieval tables from keyword queries to natural language queries. This work reports a 13\% improvement over \citep{Li:2019:TNW}.  It is worth noting that these methods are not reported using the same experimental settings as the baselines.
%} \todo{TODO(SZ): We should say sg. about performance. First, state (if correct) that all the above works employ the models from~\citep{Zhang:2018:AHT} as baselines, and report on performance improvements in the X-Y\% range in terms of NDCG@X.}

%We have discussed the line of approaches~\citep{Lehmberg:2015:MSJ, Ahmadov:2015:THI, DasSarma:2012:FRT, Yakout:2012:IEA, Nguyen:2015:RSS, Limaye:2010:ASW} that can use a table as a query.
Table-based search may be conducted for different purposes: (i) to be displayed as the answer and (ii) to serve as an intermediate step that feeds into other tasks like table mining or table augmentation. 
\citet{Ahmadov:2015:THI} leverage table elements like entities and headings as keyword queries to retrieve a ranked list of tables. The two ranked lists of tables is merged later by performing table matching to have a more complete candidate set.
Table matching is performed by dividing tables into various elements, such as table entities, headings, and columns, then computing element-level similarity.
The Mannheim Search Join Engine~\citep{Lehmberg:2015:MSJ} provides table search functionality with the overall aim to extend an input table with additional attributes (i.e., columns). \citet{Lehmberg:2015:MSJ} rely mostly on table headings by comparing the heading labels between the input and candidate tables.
Their method uses exact column heading matching to filter tables that share at least one heading with the input table. Then, all candidate tables are scored against the input table.
\citet{DasSarma:2012:FRT} find related tables for extending the seed table with extra rows or columns, referred as \emph{entity complement} and \emph{schema complement} respectively.
For \emph{entity complement} tables, which aim to augment the input table with more entities as rows, they consider the relatedness between entities of the input and candidate tables. 
For \emph{schema complement} tables, which seek to extend the input tables with more attributes, they take into account the coverage of entities and the benefits of adding additional attributes.
The above methods perform table matching in an unsupervised manner.
To enrich the diversity of search results,~\citet{Nguyen:2015:RSS} design a goodness measure for table search and selection.
They match tables by considering two tables elements: heading labels and table data. 
These two similarities are combined using a linear mixture.
\citet{Yakout:2012:IEA} consider element-wise similarity across four table elements: table data, column values, page title, and column headings. These element-wise similarities are combined by training a linear regression scorer.

\subsection{Table Mining}

Being a rich and structured source knowledge, tables have raised great interest for various mining tasks~\citep{Cafarella:2011:SDW,Cafarella:2008:WEP,JM:2009:HDW,Sarawagi:2014:OQQ,Venetis:2011:RST,Zhang:2013:ISM, Arvind:2015:NPI, Yin:2016:NEL, Sarawagi:2014:OQQ,Banerjee:2009:LRQ, Venetis:2011:RST, Chirigati:2016:KEU, Bhagavatula:2015:TEL, Zhang:2020:NED,Zhang:2020:WTE}. 
\citet{Embley:2006:TPA} present a survey of methods for table processing and applications, like table conversion from homogeneous or heterogeneous sources. These processing steps are key building blocks in table mining.
\citet{Munoz:2014:ULD} aim to recover Wikipedia table semantics and store them in RDF triples. 
Their method utilizes DBpedia to find pre-existing relations between entities in the Wikipedia tables. It then queries the DBpedia knowledge base for existing facts that involve those entities.
The prior relations contribute to extrapolate this to the rest of the table. In the end, 7.9M unique and novel RDF triples are extracted.
Similar work is taken in \citep{Cafarella:2008:WEP} based on tables extracted from a Web crawl. Instead of mining an entire corpus of tables, a single table may already store many facts, which could be answers for questions. 
\citet{Yin:2016:NEL} take a single table as a knowledge base and perform querying on it using deep neural networks, named Neural Enquirer. Neural Enquirer is fully neural system which generates distributional representations of the query and the knowledge base. It can additionally execute compositional queries as a series of operations. The training can be done in an end-to-end fashion or carried out using step-by-step supervision. 
The knowledge extracted from tables could be used to augment an existing knowledge base~\citep{Sekhavat:2014:KBA, Dong:2014:KVW}. For instance, \citet{Sekhavat:2014:KBA} design probabilistic methods to utilize table information for augmenting an existing knowledge base.  
They collect sentences containing pairs of entities in the same row by taking the tabular mentions as keyword queries. Patterns are extracted from these sentences by leveraging existing knowledge bases. Lastly, they estimate the probability of possible relations that can be added to the knowledge repository.
Recently, \citet{Zhang:2020:GCS} propose to mine new categories based on the entity sets in tables.

Another line of work concerns table annotation and classification. By mining column content, \citet{Zwicklbauer:2013:TDW} propose a method to annotate table headers with semantic type information based on the column's cells. 
The annotation is performed in three steps. First, it uses a search-based disambiguation method to annotate cell entities. Then, it resolves entity-type by retrieving a set of types for the entity candidates. Lastly, the type that occurs most frequently in the set of all types of all cells is assigned to the table header.
Studying a large number of tables in \citep{Crestan:2011:WTC}, a well defined table type taxonomy is provided for classifying HTML tables. 
They introduce a supervised framework for classifying HTML tables into a taxonomy by examining the contents of a large number of tables. 
Three types of features are considered: global layout features, layout features, and content features. 
Global layout features include the maximum number of rows, column, and cell content length. 
Layout features are solely based on the size of the cells and variance, e.g., the average length of cells.
Content features focus on cell content, including HTML features (e.g., the ratio of cells containing header) and lexical features (e.g., the ratio of cells where the content is a number).
Apart from the above mentioned problems, table mining can also include tasks like \emph{table interpretation}~\citep{Cafarella:2008:WEP, Munoz:2014:ULD, Venetis:2011:RST} and \emph{table recognition}~\citep{Crestan:2011:WTC,Zwicklbauer:2013:TDW}.  In the problem space of table mining, table search is an essential component. 

\subsection{Table Augmentation}

\emph{Table augmentation} is the task of extending a table with additional elements, e.g., new columns or rows. %~\citep{DasSarma:2012:FRT,Cafarella:2009:DIR,Lehmberg:2015:MSJ,Yakout:2012:IEA,Bhagavatula:2013:MEM}. 
Methods for extending a table with additional columns need to  
capture relevant data (i.e., existing columns), which is done with the help of table search~\citep{Lehmberg:2015:MSJ,Bhagavatula:2013:MEM,Yakout:2012:IEA}. For example, the Mannheim Search Join Engine~\citep{Lehmberg:2015:MSJ} searches the top-$k$ candidate tables from a corpus of web tables and picks relevant columns to merge. 
Extending a table with more rows also needs table retrieval~\citep{DasSarma:2012:FRT,Yakout:2012:IEA,Zhang:2017:ESA, Zhang:2017:ESA}. 
\citet{DasSarma:2012:FRT} find \emph{entity complement} tables to find the additional entities that can be put into the input table as the next rows. 
However, it stops at the table search step and only utilizes the tables.
In~\citep{Zhang:2017:ESA}, two tasks of row population and column population are proposed to extend an entity-focused table with additional rows and columns. 
The authors utilize both tables and a knowledge base for row extension. 
Specifically, their method first finds candidates from the related tables and knowledge base. 
Then, it ranks the candidates based on the similarity to (i) other entities in the tables, (ii) column headings, and (iii) the caption of the table.
Column extension relies only on tables, and it follows a similar approach. 
They find that a knowledge base and a table corpus can complement each other for table augmentation.
In recent work, \citet{Li:2019:TNW} used Word2vec to train embeddings for these two tasks and achieved state-of-the-art results.

\emph{Table completion} is the task of filling in empty cells within a table. \citet{Ahmadov:2015:THI} introduce a method to extract table values from related tables and/or to predict them using machine learning methods.
In recent work, \citet{Zhang:2019:ADC} present the CellAutoComplete system to address shortcomings of previous approaches.  Specifically, CellAutoComplete enables a table cell to have multiple, possibly conflicting values, providing the values with evidence found from other tables or the knowledge base, or predict if a cell should be left empty. Their method leverages a corpus of Wikipedia tables and a knowledge base (DBpedia) as data sources. 

\subsection{Neural Embeddings for IR}
Recently, unsupervised representation learning methods have been proposed for obtaining embeddings that predict a distributional context, i.e., word embeddings~\citep{Mikolov:2013:DRW,Pennington:2014:GGV} or graph embeddings~\citep{Perozzi:2014:DOL,Tang:2015:LLI,Ristoski:2016:RGE}.  
Such vector representations have been utilized successfully in a range of IR tasks, including ad hoc retrieval~\citep{Ganguly:2015:WEB, Bhaskar:2016:DES}, contextual suggestion~\citep{Jarana:2016:MUP}, cross-lingual IR~\citep{Vulic:2015:MCI}, community question answering~\citep{zhou-EtAl:2015:ACL-IJCNLP1}, short text similarity~\citep{Kenter:2015:STS}, and sponsored search~\citep{Grbovic:2015:CCE}.
For example, \citet{Ganguly:2015:WEB} construct a generalized language model by making use of the vector embeddings to derive the transformation probabilities between words for enhancing ad hoc retrieval effectiveness.
\citet{Bhaskar:2016:DES} propose the Dual Embedding Space Mode that uses the neural word embeddings to gauge a documents' relatedness to the query in ad hot retrieval. It exploits a novel use of both the input and output embeddings of the CBOW model to capture the topic-based semantic relationship.
\citet{Jarana:2016:MUP} exploit word embeddings to infer the vector-space representations of venues, user preferences, and users' contexture preferences for contextual suggestions.  %The evaluation demonstrate the significant effectiveness on this task.
\citet{zhou-EtAl:2015:ACL-IJCNLP1} propose to learn continuous word embeddings with metadata of category information within community question answering to fill the lexical gap. % This method demonstrates the effectiveness of these embeddings by outperforming the translation and topic-based models for question retrieval.
\citet{Kenter:2015:STS} verify the idea of investigating only using semantic features for computing short text similarity~\citep{Kenter:2015:STS} from word-level to text-level.
Most recently, transformer-based models, such as BERT~\citep{Devlin:2019:Bert}, have shown great improvements for many NLP tasks.  In this work, however, we limit ourselves to traditional (non-contextual) embeddings.

\section{Table Retrieval}
\label{sec:tableret}

In this section, we formalize the table retrieval task, and explain what information is associated with a table.

\subsection{Problem Statement}

\emph{Table search} is the task of returning a ranked list of tables from a collection of tables, in response to a query.
The relevance of each returned table $T$ is assessed independently of all other returned tables.
Tables are then sorted in descending order of their scores.
We consider two types of queries in this work.
First, when the input is a keyword query $q$, we refer to this task as \emph{keyword-based search}. 
The ranking of tables boils down to the problem of assigning a score to each table in the corpus, $\mathit{score}(q,T)$.  
Second, when the input is a table $\tilde{T}$, we term this task as \emph{table-based search}. 
The objective is to compute the similarity between an input table $\tilde{T}$ and a candidate table $T$, expressed as $\mathit{score}(\tilde{T},T)$. 

\subsection{The Anatomy of a Table} % What is in a Table?
\label{sec:tableret:whatis}

\begin{figure}[t]
   \centering
   \includegraphics[width=0.6\textwidth]{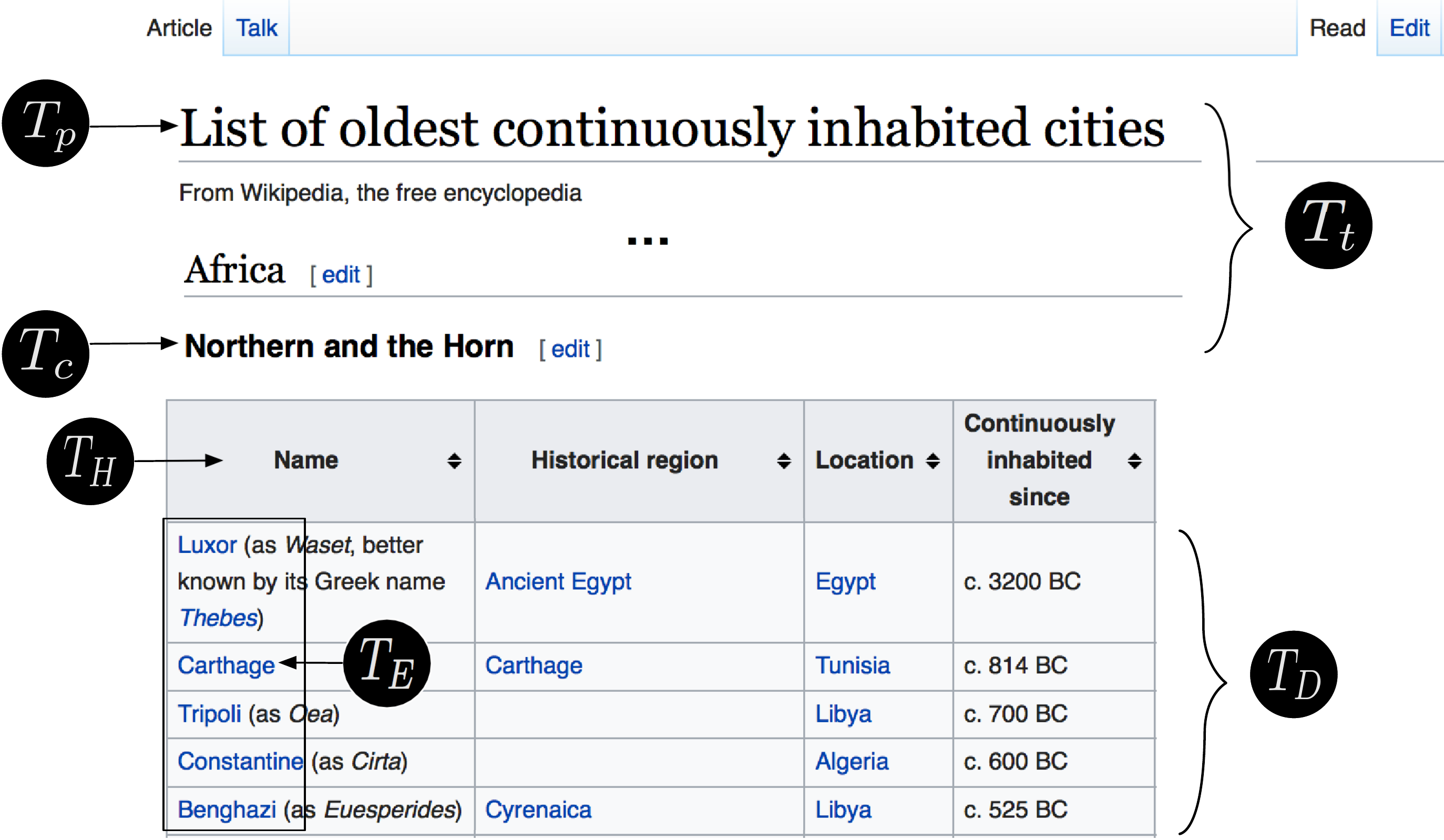}    
   \caption{Table embedded in a Wikipedia page.} % Normally, page title is the wikipedia page's name, table is equipped with section title or table caption. Also, table is usually surrounded by text.}
   \label{fig:wikitable}
\end{figure}

We shall assume a corpus of Web tables, where tables are embedded in webpages.  This means that in addition to the core table content (caption, column headings, and table body), some contextual information is also available, such as the embedding page's title.
See Table~\ref{tbl:notation} for an overview, with a corresponding illustration in Figure~\ref{fig:wikitable}.

\begin{table}[t]
  \centering
  \caption{Notation used for elements of table $T$.}
  \begin{tabular}{lll}
    \toprule
    \textbf{Symbol} & \textbf{Table element} & \textbf{Explanation} \\
    \midrule
    $T_c$ & Table caption & A brief explanation of a table \\
    $T_p$ & Page title & Title of the page where the table was extracted from \\
    $T_E$ & Table entities & Set of entities in the table \\
    $T_{E'}$ & Core column entities & Set of entities in the core column \\
    $T_H$ & Column headings & Set of table column headings\\  % was terms before
    $T_D$ & Table data & Table data, excluding column headings$^\dag$ \\
    $T_t$ & Table topic & The subject of a table \\
    \bottomrule
    \multicolumn{3}{l}{\footnotesize{$^\dag$ Could also denote, depending on the algorithm, a subset of the table (i.e., selected rows)}} \\
    \multicolumn{3}{l}{\footnotesize{ or columns)}} \\
  \end{tabular}
  \label{tbl:notation}
\end{table}

\section{Baseline Methods}
\label{sec:baselines}

Table retrieval can be performed by utilizing either keyword-based or feature-based methods.
In this section, we discuss these two directions, and detail how existing work may be applied to our tasks, depending on the input query (keywords or table).

\subsection{Keyword-based Methods}
\label{sec:tbr}

An easy and straightforward way to perform table retrieval is by treating the query as a set of keywords and adopting standard document ranking methods. 

%\subsubsection{Baselines for Keyword-based Search}

In prior work, \citet{Cafarella:2009:DIR, Cafarella:2008:WEP} utilize web search engines to retrieve relevant documents; tables are then extracted from the highest-ranked documents.  
Rather than relying on external services, we represent tables as either single- or multi-field documents and apply standard documents retrieval techniques. 

\paragraph{Single-field Document Representation}
In the simplest case, all text associated with a given table may be used as the table's representation.  This representation is then scored using existing retrieval methods, such as language models.

\paragraph{Multi-field Document Representation}
Rather than collapsing all textual content into a single-field document, it may be organized into multiple fields, such as table caption, table headers, table body, etc. (cf. Sect.~\ref{sec:tableret:whatis}).
For multi-field ranking, \citet{Pimplikar:2012:ATQ} employ a \emph{late fusion} strategy~\citep{Shuo:2017:DPF}.  That is, each field is scored independently against the query, then a weighted sum of the field-level similarity scores is taken:
\begin{equation}
	\mathit{score}(q,T) = \sum_{i} w_i \times \mathit{score}(q, f_i) ~,
\end{equation}
where $f_i$ denotes the $i$th (document) field for table $T$ and $w_i$ is the corresponding field weight (such that $\sum_i w_i = 1$). Field-level similarity, $\mathit{score}(q, f_i)$, may be computed using any standard retrieval method.  We use language models in our experiments.

%\subsubsection{Baselines for Table-based Search}
%\paragraph{Keyword-based Search}
%\label{sec:existing:keyword}

%\citet{Ahmadov:2015:THI} use table entities and table headings as queries.  Additionally, in this paper we also consider using the table caption as a query.  
%Formally, we let $d_{T}$ denote the document-based representation of table $T$.  The candidate table's score is computed by taking the terms from $\tilde{T}_E$, $\tilde{T}_H$, or $\tilde{T}_c$ as the keyword query $q$: $\mathit{score}(\tilde{T},T) = \mathit{score}(q, d_T)$.

\subsection{Feature-based Methods for Keyword-based Search}  
\label{sec:fbr}

Another line of work uses feature-based methods.
The idea is to represent both the query and the candidate table as a feature vector, and obtain a retrieval function using supervised learning~\citep{Liu:2011:LRI}.  In a standard document retrieval setting, features are commonly categorized into three groups: (i) document, (ii) query, and (iii) query-document features~\citep{Qin:2010:LBC}.  Analogously, we distinguish between three types of features: (i) table, (ii) query, and (iii) query-table features.
In Table~\ref{tbl:features}, we summarize the features from previous work on table search~\citep{Cafarella:2008:WEP,Bhagavatula:2013:MEM}.  We also include a number of additional features that have been used in other retrieval tasks, such as document and entity ranking. %; we do not regard these as novel contributions.

\begin{table}[!t]
\centering
\caption{Baseline features for keyword-based search (LTR-k).}
\footnotesize
\begin{tabular}{p{2cm}p{7cm}p{4cm}p{2cm}}
	\toprule
	\multicolumn{2}{l}{\textbf{Query features}} & \textbf{Source} & \textbf{Value} \\
	\midrule
	QLEN & Number of query terms & \citep{Tyree:2011:PBR} &  \{1,...,n\} \\
	$\mathrm{IDF}_f$ & Sum of query IDF scores in field $f$ & \citep{Qin:2010:LBC} & $[0,\infty)$ \\	
	\midrule
	\multicolumn{4}{l}{\textbf{Table features}} \\
	\midrule
	\#rows & Number of rows in the table & \citep{Cafarella:2008:WEP,Bhagavatula:2013:MEM} & \{1,...,n\} \\
	\#cols & Number of columns in the table & \citep{Cafarella:2008:WEP,Bhagavatula:2013:MEM} & \{1,...,n\} \\
	\#NULLs & Number of empty table cells & \citep{Cafarella:2008:WEP,Bhagavatula:2013:MEM} &  \{0,...,n\} \\
	PMI & ACSDb-based schema coherency score & \citep{Cafarella:2008:WEP} & $(-\infty,\infty)$ \\	
	inLinks & In-link count of the embedding page & \citep{Bhagavatula:2013:MEM} & \{0,...,n\} \\
	outLinks & Out-link count of the embedding page & \citep{Bhagavatula:2013:MEM} & \{0,...,n\}\\
	pageViews & Page view count of the embedding page & \citep{Bhagavatula:2013:MEM} & \{0,...,n\}\\
	tableImportance & Inverse of number of tables on the page & \citep{Bhagavatula:2013:MEM} & $(0,1]$ \\
	tablePageFraction & Table size to page size ratio & \citep{Bhagavatula:2013:MEM} & $(0,1]$ \\
	\midrule
	\multicolumn{4}{l}{\textbf{Query-table features}} \\
	\midrule
	\#hitsLC & Total query term frequency in leftmost column & \citep{Cafarella:2008:WEP} & \{0,...,n\} \\
	\#hitsSLC & Total query term frequency in second-to-leftmost column & \citep{Cafarella:2008:WEP} & \{0,...,n\} \\
	\#hitsB & Total query term frequency in table body & \citep{Cafarella:2008:WEP} & \{0,...,n\} \\
	qInPgTitle & Ratio of the number of query tokens found in page title to total number of tokens & \citep{Bhagavatula:2013:MEM} & $[0,1]$ \\
	qInTableTitle & Ratio of the number of query tokens found in table title to total number of tokens & \citep{Bhagavatula:2013:MEM} & $[0,1]$ \\
	yRank & Rank of the table's Wikipedia page in web search results for the query & \citep{Bhagavatula:2013:MEM} & \{1,...,n\} \\
	MLM similarity & Language modeling score between query and multi-field document representation of the table & \citep{Chen:2016:ESL} & ($-\infty$,0) \\
		\bottomrule
\end{tabular}
\label{tbl:features}
\end{table}

\subsubsection{Query Features}
Query features have been shown to improve retrieval performance for document ranking~\citep{Macdonald:2012:UQF}.  
We adopt two query features from document retrieval, namely, 
the number of terms in the query~\citep{Tyree:2011:PBR}, 
and query IDF~\citep{Qin:2010:LBC} according to:
$\mathit{IDF}_f(q) = \sum_{t \in q} \mathit{IDF}_f(t)$,
where $\mathit{IDF}_f(t)$ is the IDF score of term $t$ in field $f$.  This feature is computed for the following fields: page title, section title, table caption, table heading, table body, and ``catch-all'' (the concatenation of all textual content in the table).
%Note that this group of features is only used in keyword-based search.

\subsubsection{Table Features}
\label{subsec:tf}

Table features depend only on the table itself and aim to reflect the quality of the given table (irrespective of the query).
Some features are simple characteristics, like the number of rows, columns, and empty cells~\citep{Cafarella:2008:WEP,Bhagavatula:2013:MEM}.  
One important feature is Point-wise Mutual Information (PMI), which is taken from linguistics research, and expresses the coherency of a table.  The correlation between two table headings cells, $h_i$ and $h_j$, is given by: $PMI(h_i,h_j) = \log \big(P(h_i,h_j) / (P(h_i)P(h_j))\big)$.
For example, ``address'' and ``name'' are more likely to co-occur as column headings than ``address'' and ``wins.''
A table's PMI is computed by calculating the PMI values between all pairs of column headings of that table, and then taking their average. 
Following~\citep{Cafarella:2008:WEP}, we compute PMI by obtaining frequency statistics from the Attribute Correlation Statistics Database (ACSDb)~\citep{Cafarella:2008:URW}, which contains table heading information derived from millions of tables extracted from a large web crawl.

%Tables are typically embedded in (web) pages.  
Another group of features are derived from the webpage that embeds the table, by considering its connectivity (inLinks and outLinks), popularity (pageViews), and the table's importance within the page (tableImportance and tablePageFraction).

\subsubsection{Query-Table Features}
\label{sec:existing}

Features in the last group express the degree of matching between the keyword query and a given candidate table.  
%For keyword-table search,
This matching may be based on occurrences of query terms in the page title (qInPgTitle) or in the table caption (qInTableTitle).
Alternatively, it may be based on specific parts of the table, such as the leftmost column (\#hitsLC), second-to-left column (\#hitsSLC), or table body (\#hitsB).
The rank at which a table's parent page is retrieved by an external search engine is also used as a feature (yRank).  (In our experiments, we use the Wikipedia search API to obtain this ranking.)
Furthermore, we take the Mixture of Language Models (MLM) similarity score~\citep{Ogilvie:2003:MLM} as a feature, which is actually the best performing method among the four text-based baseline methods (cf. Sect.~\ref{sec:eval}).

We utilize table features, query features, and query-table features to train a learn-to-rank scorer, to serve as a strong baseline, and label it \emph{LTR-k}.
Specifically, we manually label the candidates retrieved by the unsupervised methods introduced in Sect.~\ref{sec:tbr} as training data (cf. Sect.~\ref{sec:testcoll:tablecorpus}), and combine different set of features for \emph{LTR-k} to train the scorers (cf. Sect.\ref{sec:eval:baselines}).
Importantly, all these features are based on lexical matching.  Our goal in this paper is to also enable semantic matching; this is what we shall discuss in Section~\ref{sec:sem}. \\

\subsection{Feature-based Methods for Table-based Search}  
\label{sec:fmt}

%\subsubsection{Baselines for Table-based Search}
%\label{sec:existing}

\begin{table}[t]
  \centering
  \caption{Table elements used in existing table-based methods.}
  \begin{tabular}{lc@{~~~~~}c@{~~~~~}c@{~~~~~}c@{~~~~~}c}
    \toprule
    \textbf{Method}                                                 & \textbf{$T_c$}    & \textbf{$T_p$}  & \textbf{$T_E$} & \textbf{$T_H$} & \textbf{$T_D$} \\
    \midrule
    Keyword-based search using $T_E$ &         &         & $\surd$ &         &         \\
    Keyword-based search using $T_H$  &         &         &         & $\surd$ &         \\
    Keyword-based search using $T_c$  & $\surd$ &         &         &         &         \\
    Mannheim Search Join Engine          &         &         &         & $\surd$ &         \\
    Schema complement                      &         &         & $\surd$ & $\surd$ &         \\
    Entity complement                      &         &         & $\surd$ &         &         \\
    Nguyen et al.                         &         &         &         & $\surd$ & $\surd$ \\
    InfoGather                   &         & $\surd$ &         & $\surd$ & $\surd$ \\
    \bottomrule
  \end{tabular}
  \label{tbl:baseline_method1}
\end{table}
For table-based search, table features can still be employed (cf. the middle block in Table~\ref{tbl:baseline_method1}).  In fact, they may be computed for both the input and candidate tables.  (When computed for the input table, these essentially become the ``query features.'')
Query-table features, however, are different from those in keyword-based search, as we need to perform table-to-table matching as opposed to keyword-to-table matching. 

We present a number of existing methods from the literature that can be used to perform table matching.  
On the high level, all these methods operate by (i) subdividing tables into a number of \emph{table elements} (such as page title ($T_p$), table caption ($T_c$), table topic ($T_t$), column headings ($T_H$), table entities ($T_E$), and table data ($T_D$)), (ii) measuring the similarity between various elements of the input and candidate tables, and (iii) in case multiple elements are considered, combining these element-level similarities into a final score.  Table~\ref{tbl:baseline_method1} provides an overview of existing methods and the table elements they utilize.

\paragraph{Mannheim Search Join Engine}
\label{sec:existing:msje}

The Mannheim Search Join Engine (MSJE)~\citep{Lehmberg:2015:MSJ} provides table search functionality with the overall aim to extend an input table with additional attributes (i.e., columns).  
First, it uses exact column heading matching to filter tables that share at least one heading with the input table: $\mathcal{T} = \{T : |\tilde{T}_H \cap T_H| > 0\}$.
Then, all tables in $\mathcal{T}$ are scored against the input table using the FastJoin matcher~\citep{Wang:2011:FEM}. Specifically, \citet{Lehmberg:2015:MSJ} adapt edit distance with a threshold of $\delta$ to measure the similarity between the input and candidate tables' heading terms, $w(t_i,t_j)$, where $t_i \in \tilde{T}_H$ and $t_j \in T_H$. 
Terms in $\tilde{T}_H$ and $T_H$ form a bipartite graph, with $w(t_i,t_j)$ as edge weights. 
Let $|\tilde{T}_H\tilde{\cap}_{\delta}T_H|$ denote the \emph{maximum weighted bipartite matching score} on the graph's adjacency matrix, considering edges whose weight exceeds the edit distance threshold $\delta$.  
Then, the Jaccard similarity of two tables is expressed as:
\begin{equation*}
	\mathit{score}(\tilde{T},T) = \frac{|\tilde{T}_H\tilde{\cap}_{\delta}T_H|}{|\{t: t \in \tilde{T}_H\}|+|\{t: t \in T_H\}|-|\tilde{T}_H\tilde{\cap}_{\delta}T_H|} ~,
\end{equation*}
where $|\{t: t \in T_H\}|$ denotes the number of unique terms in the column headings of $T$.

\paragraph{Schema Complement}
\label{sec:existing:sc}

\citet{DasSarma:2012:FRT} search for related tables with the overall goal of extending the input table with additional attributes (referred to as \emph{schema complement} in~\citep{DasSarma:2012:FRT}).  
For this task, they consider two factors: (i) the coverage of entities and (ii) the benefits of adding additional attributes.  The final matching score is computed as:
\begin{equation}
	\mathit{score}(\tilde{T}, T) = S_{EC}(\tilde{T}, T) \times S_{HB}(\tilde{T}, T) . \label{eq:schemacompl}
\end{equation}
The first component, entity coverage (EC), computes the entity overlap between two tables:
\begin{equation}
	S_{EC}(\tilde{T}, T)=\frac{|\tilde{T}_E \cap T_E|}{|\tilde{T}_E|} . \label{eq:ECover}
\end{equation}
The second component in Eq.~\eqref{eq:schemacompl} estimates the benefit of adding an additional column heading $h$ to the input table:
\begin{equation*}
	HB(\tilde{T}_H,h) = \frac{1}{|\tilde{T}_H|}\sum_{\tilde{h} \in \tilde{T}_H}\frac{\#(\tilde{h},h)}{\# (\tilde{h})}~,
%\label{eq:sb}
\end{equation*}
where $\#(\tilde{h},h)$ is number of tables containing both $\tilde{h}$ and $h$ as column headings, and $\#(\tilde{h})$ is the number of tables containing $\tilde{h}$.  The heading benefit between two tables, $S_{HB}(\tilde{T}, T)$, is computed by aggregating the benefits of adding all headings $h$ from $T$ to $\tilde{T}$:
\begin{equation*}
	S_{HB}(\tilde{T}, T) = \mathit{aggr}(HB(\tilde{T}_H,h)) ~.
\end{equation*}
The aggregation function $\mathit{aggr}()$ can be sum, average, or max.

\paragraph{Entity Complement}
\label{sec:existing:ec}

In addition to schema complement tables, \citet{DasSarma:2012:FRT} also search for \emph{entity complement} tables, in order to augment the input table with additional entities (as rows).  This method considers the relatedness between entities of the two tables:
\begin{equation*}
	\mathit{score}(\tilde{T}, T) = \frac{1}{|\tilde{T}_E||T_E|}\sum_{\tilde{e} \in \tilde{T}_E} \sum_{e \in T_E} \mathit{sim}(\tilde{e}, e) ,
	\label{eq:entity_set}
\end{equation*}
where $\mathit{sim}(\tilde{e}, e)$ is a pairwise entity similarity measure.  Specifically, we employ the \emph{Wikipedia Link-based Measure} (WLM)~\citep{Milne:2008:ELM}, which estimates the semantic relatedness between two entities based on other entities they link to:
\begin{equation*}
	\mathit{sim}_{WLM}(e,\tilde{e})=1-\frac{\log(\max(|\mathcal{L}_e|,|\mathcal{L}_{\tilde{e}}|))-\log(|\mathcal{L}_e \cap \mathcal{L}_{\tilde{e}}|)}{\log(|\mathcal{E}|-\log(\min(|\mathcal{L}_e|,|\mathcal{L}_{\tilde{e}}|)))} ~,
%\label{eq:wlm}
\end{equation*}
where $\mathcal{L}_e$ is the set of outgoing links of $e$ (i.e., entities $e$ links to) and $|\mathcal{E}|$ is the total number of entities in the knowledge base.

\paragraph{Nguyen et al.} %Diversified Table Search
\label{sec:existing:nguyen}

\citet{Nguyen:2015:RSS} match tables by considering both their headings and content (table data).  These two similarities are combined using a linear mixture:
\begin{equation*}
	\mathit{score}(\tilde{T}, T) = \alpha \times \mathit{sim}_H(\tilde{T}, T) + (1-\alpha) \times \mathit{sim}_D(\tilde{T}, T) ~.
\end{equation*}
The heading similarity $\mathit{sim}_H$ is computed by first creating a similarity matrix between the heading terms of $\tilde{T}_H$ and $T_H$, as in Sect.~\ref{sec:existing:msje}.  
Specifically, for two sets of headings $\tilde{T}_H$ and $T_H$, it first constructs a $|\tilde{T}_H| \times |T_H|$ similarity matrix $m(\tilde{T}_H, T_H)$, where $m_{ij}(\tilde{T}_H, T_H)$ denotes the degree of similarity between the heading $i$ of $\tilde{T}_H$ and $j$ of $T_H$, respectively. 
Next, an attribute correspondence subgraph $C \subseteq (|\tilde{T}_H| \times |T_H|)$ is obtained by solving the \emph{maximum weighted bipartite sub-graph problem}~\citep{Anan:2007:OSM}. 
Taking column headings as vertices of a graph and $m_{ij}(\tilde{T}_H, T_H)$ as edge weights, the task is to find the maximum weight bipartite subgraph, such that the vertices of the subgraph can be divided into two disjoint and independents set $U$ and $V$, where every edge connects a vertex in $V$ to $U$.
Finally, heading similarity is computed as:
\begin{equation*}
	\mathit{sim}_H(\tilde{T}, T) = \frac{\sum_{(i,j) \in C}w_{t_i,t_j}(\tilde{T}_H, T_H)}{\max (|\tilde{T}_H|,|T_H|)} ~.
\end{equation*}
Data similarity is measured based on columns.  Each table column is represented as a binary term vector, $\mathbf{T}_{D,i}$, where each element indicates the presence ($1$) or absence ($0$) of a given term in column $i$ of table $T$. 
The similarity between two columns is measured by their cosine similarity. %, cf. Eq.~\eqref{eq:cosine}.
Table similarity considers all column combinations of $\tilde{T}$ and $T$.  To account for the high number of possible combinations, for each table column, only the most similar column is considered from the other table:
\begin{equation*}	
	\mathit{sim}_D(\tilde{T}, T) = \\
	\frac{1}{2}\big(\sum_{i} \max_{j} \cos(\tilde{\mathbf{T}}_{D,i},\mathbf{T}_{D,j}) + 
	\sum_{j} \max_{i} \cos(\tilde{\mathbf{T}}_{D,i},\mathbf{T}_{D,j}) \big) ~.
\end{equation*}

\paragraph{InfoGather}
\label{sec:existing:infogather}

Following \citet{Yakout:2012:IEA}, we consider element-wise similarity across four table elements: table data, column values, page title, and column headings.
%\footnote{We note that \citet{Yakout:2012:IEA} consider additional signals as well, such as context and URL similarity.  We do not include those because \todo{XXX}.} 
Element-wise similarities are combined by training a linear regression scorer:
\begin{equation*}
	\mathit{score}(\tilde{T}, T) = \sum_{x} w_x \times \mathit{sim}_x(\tilde{T},T) ~,
\end{equation*}
where $x$ is a given table element, $\mathit{sim}_x()$ is the element-level similarity score, and $w_x$ is the weight (importance) of that element.
Each table element is expressed as a term vector, denoted as $\tilde{\mathbf{T}}_x$ and $\mathbf{T}_x$ for element $x$ of the input and candidate tables, respectively.  Element-level similarity is then estimated using the cosine similarity between the two term vectors:
\begin{equation}
	\mathit{sim}_x(\tilde{T}, T) = \cos(\tilde{\mathbf{T}}_x, \mathbf{T}_x) = \frac{\tilde{\mathbf{T}}_x \cdot \mathbf{T}_x}{||\tilde{\mathbf{T}}_x|| \times ||\mathbf{T}_x||} ~.
	\label{eq:cosine}
\end{equation}
Specifically, following~\citep{Yakout:2012:IEA}, for the table data and page title elements we use IDF weighting, while for column heading and column values, we employ TF-IDF weighting. 
% on the respective fields.

\paragraph{Our baselines}

%We first take the existing table matching methods as our baselines.
%Apart from them,
We leverage all the table matching scores introduced in this section as features in a learning-to-rank scorer. 
It serves as the first strong table-based search baseline and is labeled \emph{LTR-t1}. 
Additionally, we also incorporate table features (cf. Sect.~\ref{subsec:tf}), computed for both the input and candidate tables, in a second strong baseline \emph{LTR-t2}.

\section{Semantic Matching}
\label{sec:sem}

\begin{figure*}[t]
   \centering
   \includegraphics[width=0.9\textwidth]{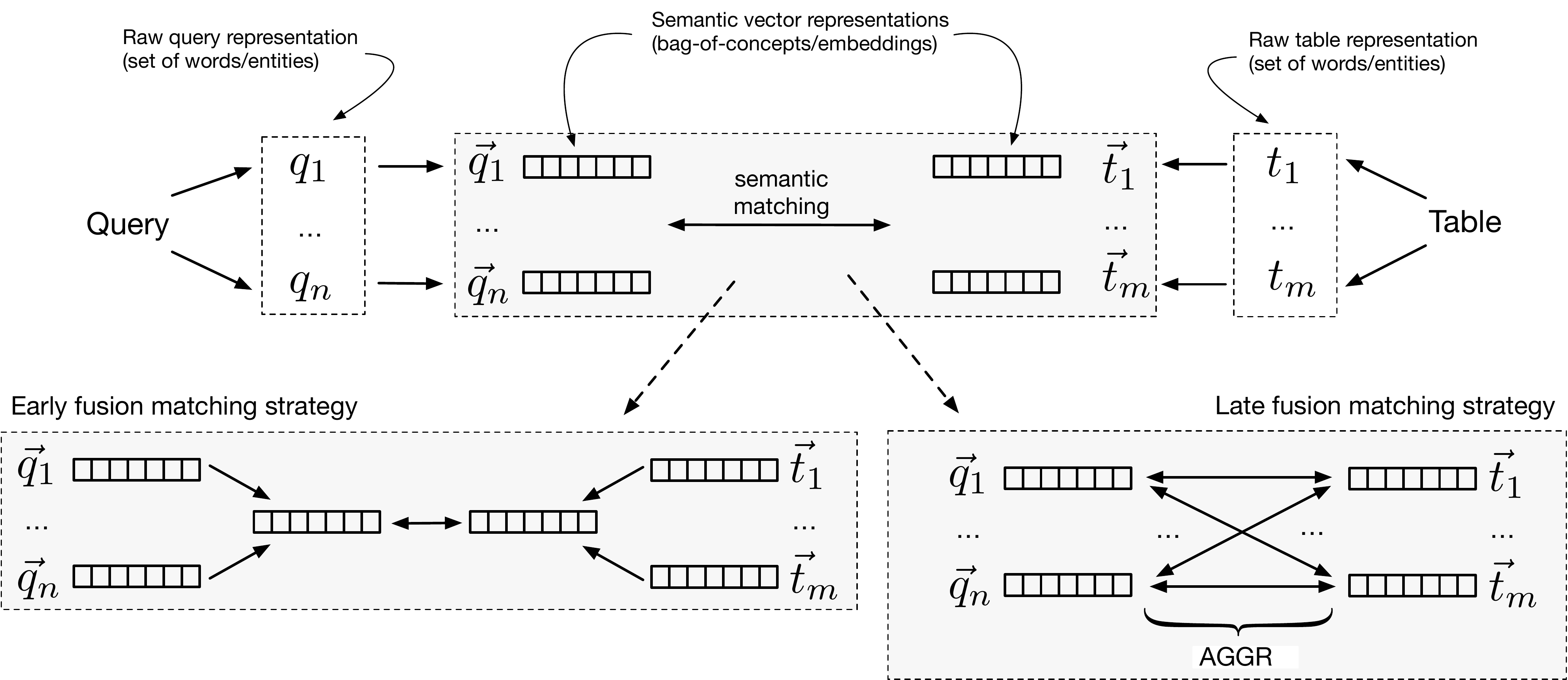} 
   \caption{Our methods for computing query-table similarity for keyword-based search using semantic representations. }
\label{fig:e3v}
\end{figure*}
This section presents our main contribution, which is a framework for performing semantic matching for table retrieval.  %a set of novel semantic matching methods for table retrieval.  
The main idea is to go beyond lexical matching by representing both queries and tables in some semantic space, and measuring the similarity of those semantic (vector) representations.
Our approach consists of three main steps, which are illustrated in Figure~\ref{fig:e3v}.  These are as follows (moving from outwards to inwards on the figure):
\begin{enumerate}
	\item The ``raw'' content of a query/table is represented as a set of terms, where terms can be either words or entities (Sect.~\ref{sec:sem:repr_raw}).
	\item Each of the raw terms is mapped to a semantic vector representation (Sect.~\ref{sec:sem:repr_sem}).
	\item The semantic similarity (matching score) between a query-table pair is computed based on their semantic vector representations (Sect.~\ref{sec:sem:match}).
\end{enumerate}
We compute query-table similarity using all possible combinations of semantic representations and similarity measures, and use the resulting semantic similarity scores as features in a learning-to-rank approach.  
Table~\ref{tbl:features2} summarizes these features.

\begin{table}[t]
\centering
\caption{Semantic similarity features.
Each row represents 4 features (one for each similarity matching method, cf. Table~\ref{tbl:sim_measures}).  All features are in $[-1,1]$.}
%\vspace*{-0.5\baselineskip}
\begin{tabular}{p{1.5cm}p{4.5cm}p{2.4cm}}
	\toprule
	\textbf{Features} & \textbf{Semantic repr.} & \textbf{Raw repr.} \\
	\midrule
		Entity\_* & Bag-of-entities & entities \\ % & Entity\_Early, Entity\_Late\_\{max$|$avg$|$sum\} \\
%		Category\_* & Bag-of-categories & entities \\ %& Category\_Early, Category\_Late\_\{max$|$avg$|$sum\} \\
		Word\_* & Word embeddings & words \\ % & Word\_Early, Word\_Late\_\{max$|$avg$|$sum\} \\
		Graph\_* & Graph embeddings & entities \\ % & Graph\_Early, Graph\_Late\_\{max$|$avg$|$sum\} \\	
	\bottomrule
\end{tabular}
%\vspace*{-0.5\baselineskip}
\label{tbl:features2}
\end{table}

\subsection{Content Extraction} %Raw Content Representation}
\label{sec:sem:repr_raw}

We represent the ``raw'' content of the query/table as a set of terms, where terms can be either words (string tokens) or entities (from a  knowledge base).  We denote these as $\{q_1, \dots, q_n\}$ and $\{t_1, \dots, t_m\}$ for query $q$ and table $T$, respectively.  
%Below, we describe how to extract content using words and entities. 

\subsubsection{Word-based}
It is a natural choice to simply use word tokens to represent query/table content.  That is, $\{q_1, \dots, q_n\}$ is comprised of the unique words in the query. 
As for the table, we let $\{t_1, \dots, t_m\}$ contain all unique words from the title, caption, and headings of the table.
Mind that at this stage we are only considering the presence/absence of words.  During the query-table similarity matching, the importance of the words will also be taken into account (Sect.~\ref{sec:sem:match_early}).

\subsubsection{Entity-based}
Many tables are focused on specific entities~\citep{Zhang:2017:ESA}.  Therefore, considering the entities contained in a table amounts to a meaningful   representation of its content. 
We use the DBpedia knowledge base as our entity repository. 
Since we work with tables extracted from Wikipedia, the entity annotations are readily available (otherwise, entity annotations could be obtained automatically, see, e.g., ~\citep{Venetis:2011:RST}).
Importantly, instead of blindly including all entities mentioned in the table, we wish to focus on salient entities.
Salient entities come from three sources: the core column, page title, and table caption.
It has been observed in prior work~\citep{Venetis:2011:RST, Bhagavatula:2015:TEL} that tables often have a \emph{core column}, containing mostly entities, while the rest of the columns contain properties of these entities (many of which are entities themselves).
We describe our core column detection method in Sect.~\ref{sec:sem:core_col}. 
In addition to the entities taken directly from the body part of the table, we also include entities that are related to the page title ($T_{p}$) and to the table caption ($T_{c}$).  We obtain those by using the page title and the table caption, respectively, to retrieve relevant entities from the knowledge base.  We write $R_k(s)$ to denote the set of top-$k$ entities retrieved for the query $s$.  We detail the entity ranking method in Sect.~\ref{sec:sem:er}.  
Finally, the table is represented as the union of three sets of entities, originating from the core column, page title, and table caption: 
$\{t_1, \dots, t_m\} = T_{E'} \cup R_k(T_{p}) \cup R_k(T_{c})$.
%The final set of entities used to represent the table is $\{t_1, \dots, t_m\} = E_{cc} \cup E_{pt} \cup E_{tc}$.

To get an entity-based representation for the query, we issue the query against a knowledge base to retrieve relevant entities, using the same retrieval method as above.  I.e., $\{q_1, \dots, q_n\} = R_k(q)$.

\subsubsection{Core Column Detection}
\label{sec:sem:core_col}

We introduce a simple and effective core column detection method.  It is based on the notion of \emph{column entity rate}, which is defined as the ratio of cells in a column that contain an entity.  We write $\mathit{cer}(T_{c[j]})$ to denote the column entity rate of column $j$ in table $T$.
Then, the index of the core column becomes: 
$\arg\max_{j=1..T_{|c|}} \mathit{cer}(T_{c[j]})$,
where $T_{|c|}$ is the number of columns in $T$.
%where $T_{c[j]}$ refers to the $j$th column of table $T$, and $\mathit{cer}$ denotes the   
%\sz{The column with highest $cer$ will be selected as core column $cc$.}

\subsubsection{Entity Retrieval}
\label{sec:sem:er}

We employ a fielded entity representation with five fields (names, categories, attributes, similar entity names, and related entity names) and rank entities using the Mixture of Language Models approach~\citep{Ogilvie:2003:MLM}.  We return the top-$k$ entities, where $k$ is set to 10.  The field weights are set uniformly. This corresponds to the MLM-all model in~\citep{Hasibi:2017:DVT} and is shown to be a solid baseline.  We acknowledge that more advanced entity retrieval methods exist~\citep{Balog:2018:EOS}, however, this is outside the focus of the current work.
%the smoothing parameter is set as 2000

\subsection{Semantic Representations}
\label{sec:sem:repr_sem}

Next, we embed the query/table terms in a semantic space. That is, we map each table term $t_i$ to a vector representation $\vec{t}_i$, where $\vec{t}_i[j]$ refers to the $j$th element of that vector.  For queries, the process goes analogously. 
We discuss two main kinds of semantic spaces, bag-of-concepts and embeddings. %, with two alternatives within each.  
The former uses sparse and discrete, while the latter employs dense and continuous-valued vectors.  A particularly nice property of our semantic matching framework is that it allows us to deal with these two different types of representations in a unified way.

\subsubsection{Bag-of-concepts}
One alternative for moving from the lexical to the semantic space is to represent tables/queries using specific concepts. In this work, we use entities from a knowledge base.
Entities have been used in the past for various retrieval tasks, in duet with the traditional bag-of-words content representation.
For example, entity-based representations have been used for document retrieval~\citep{Xiong:2017:WDR,Raviv:2016:DRU}.
One important difference from previous work is that instead of representing the entire query/table using a single semantic vector, we map each individual query/table term to a separate semantic vector, thereby obtaining a richer representation.  
%This not only results in a richer representation, but also allows us to treat the bag-of-concepts and embeddings representations in a uniform framework.

We use the entity-based raw representation from the previous section, that is, $t_i$ and $q_j$ are specific entities. Below, we explain how table terms $t_j$ are projected to $\vec{t_i}$, which is a sparse discrete vector in the entity space; for query terms it follows analogously. Each element in $\vec{t}_i$ corresponds to a unique entity. Thus, the dimensionality of $\vec{t}_i$ is the number of entities in the knowledge base (on the order of millions). $\vec{t}_i[j]$ has a value of $1$ if entities $i$ and $j$ are related (there exists a link between them in the knowledge base), and $0$ otherwise. %~\footnote{We skip bag-of-categories for this version as it has little impact on this task.}
%	\item[Bag-of-categories] Each element in $\vec{t}_i$ corresponds to a Wikipedia category.  Thus, the dimensionality of $\vec{t}_i$ amounts to the number of Wikipedia categories (on the order hundreds of thousands).  The value of $\vec{t}_i[j]$ is $1$ if entity $i$ is assigned to Wiki\-pe\-dia category $j$, and $0$ otherwise.
%\end{description}

\subsubsection{Embeddings}
\label{sec:sem:embeddings}

Embedding-based representations representations have been utilized successfully in a range of IR tasks, including ad hoc retrieval~\citep{Ganguly:2015:WEB, Bhaskar:2016:DES}, contextual suggestion~\citep{Jarana:2016:MUP}, cross-lingual IR~\citep{Vulic:2015:MCI}, community question answering~\citep{zhou-EtAl:2015:ACL-IJCNLP1}, short text similarity~\citep{Kenter:2015:STS}, and sponsored search~\citep{Grbovic:2015:CCE}.
We consider both word-based and entity-based raw representations from the previous section and use the corresponding (pre-trained) embeddings.  We shall use well-established embedding methods noting that these may be substituted with (more recent) alternatives~\citep{Pilehvar:2020:ENL}.

\begin{description}
	\item[Word embeddings] We map each query/table word to a word embedding. Specifically, we use word2vec~\citep{Mikolov:2013:DRW} with 300 dimensions, trained on Google News data.\footnote{It has been verified in \citep{Li:2019:TNW} that word embeddings trained based on Google News data and on a table corpus lead to comparable performance on the keyword-based table retrieval task.  Therefore, in the interest of simplicity, we shall utilize word embeddings trained on the former in this work, and leave embedding learning for table retrieval for the future.}
	\item[Graph embeddings] We map each query/table entity to a graph embedding.  In particular, we use RDF2vec~\citep{Ristoski:2016:RGE} with 200 dimensions, trained on DBpedia 2015-10.
\end{description}

\subsection{Similarity Measures}
\label{sec:sem:match}

The final step is concerned with the computation of the similarity between a query-table pair, based on the semantic vector representations we have obtained for them.  We introduce two main strategies, which yield four specific similarity measures. These are summarized in Table~\ref{tbl:sim_measures}.

\subsubsection{Early Fusion}
\label{sec:sem:match_early}

The first idea is to represent the query and the table each with a single vector.  Their similarity can then simply be expressed as the similarity of the corresponding vectors.  We let $\vec{C}_q$ be the centroid of the query term vectors ($\vec{C}_q = \sum_{i=1}^n \vec{q}_i/n$).
Similarly, $\vec{C}_T$ denotes the centroid of the table term vectors.
The query-table similarity is then computed by taking the cosine similarity of the centroid vectors.  Due to the compositional capabilities of embeddings, this simple centroid-based summarization of content is shown to achieve good performance~\citep{Rossiello:2017:CTS}.
%
%\begin{equation}
%	\mathrm{sim}_{\mathit{early}}(q,T) = \cos(\vec{C}_{q}, \vec{C}_{T}) ~.
%	\label{eq:cos_sim_o2v_e}
%\end{equation}
%
When query/table content is represented in terms of words, we additionally make use of word importance by employing standard TF-IDF term weighting.  Note that this only applies to word embeddings (as the other two semantic representations are based on entities).  In case of word embeddings, the centroid vectors are calculated as $\vec{C}_T = \sum_{i=1}^m \vec{t}_i \times TFIDF (t_i)$. The computation of $\vec{C}_q$ follows analogously.

\begin{table}[t]
\centering
\caption{Similarity measures.}
\begin{tabular}{p{1.5cm}p{7.5cm}}
	\toprule
	\textbf{Measure}  & \textbf{Equation} \\
	\midrule
	Early  & $\cos(\vec{C}_{q}, \vec{C}_{T})$ \\
	Late-max & $\mathrm{max}(\{\cos(\vec{q}_i,\vec{t}_j) : i \in [1..n], j \in [1..m] \})$\\
	Late-sum & $\mathrm{sum}(\{\cos(\vec{q}_i,\vec{t}_j) : i \in [1..n], j \in [1..m] \})$\\
	Late-avg & $\mathrm{avg}(\{\cos(\vec{q}_i,\vec{t}_j) : i \in [1..n], j \in [1..m] \})$\\
	\bottomrule
\end{tabular}
\label{tbl:sim_measures}
\end{table}

\subsubsection{Late Fusion}
\label{sec:sem:match_late}

Instead of combining all semantic vectors $q_i$ and $t_j$ into a single one, late fusion computes the pairwise similarity between all query and table  vectors first, and then aggregates those.  We let $S$ be a set that holds all pairwise cosine similarity scores: 
$S = \{\cos(\vec{q}_i,\vec{t}_j) : i \in [1..n], j \in [1..m] \}$.
The query-table similarity score is then computed as $\mathrm{aggr}(S)$, where $\mathrm{aggr}()$ is an aggregation function.  Specifically, we use $\max()$, $\mathrm{sum}()$ and $\mathrm{avg}()$ as aggregators; see the last three rows in Table~\ref{tbl:sim_measures} for the equations.

\subsection{Specific Instantiations}

Next, we discuss specific instantiations of our framework for keyword-based and table-based search.

\subsubsection{Keyword-based Search}

%We illustrate our methods for computing query-table similarity using semantic representations in Figure~\ref{fig:e3v}.
We compute query-table similarity using all possible combinations of semantic representations (3) and similarity measures (4), and use the resulting (3$\times$4) semantic similarity scores as features in a learning-to-rank approach.  In addition, we also leverage the full set of features in Table~\ref{tbl:features}. 
This specific instantiation of our framework is labeled as \emph{STR-k}.

\subsubsection{Table-based Search}

Existing methods have only considered matching between elements of the same type, referred to as \emph{element-wise} matching.  Our framework also enables us to measure the similarities between elements of different types in a principled way, referred to as \emph{cross-element} matching. Finally, as before, we can also utilize table features that characterize the input and candidate tables.
Below, we detail the set of features used for measuring element-level similarity.

%\todo{Refer back to 4.1.2: Whenever we compute element-level similarity, we use 4 features: }
\begin{description}
	\item [Element-wise similarity] We compute the similarity between elements of the same type from the input and candidate tables.  Each table element may be represented in up to three semantic spaces. Then, in each of those spaces, similarity is measured using the four element-level similarity measures (early, late-max, late-sum, and late-avg).  Element-wise features are summarized in the left half of Table~\ref{tbl:element-wise-feature}.
	\item [Cross-element similarity] This approach compares table elements of different types in an asymmetrical way. Each pair of elements need to be represented in the same semantic space. Then, the same element-level similarity measures may be applied, as before.  We list the cross-element similarity features in the right half of Table~\ref{tbl:element-wise-feature}.
\end{description}

\begin{table}[!tbp]
  \centering
  \caption{Element-wise and cross-element features for table-based search. The dimension is $r \times s \times m$, where $r$ is reflection (1 for element-wise and 2 for cross-element), $s$ is the number of semantic spaces, and $m$ is the number of element-wise similarity measures.}
  \begin{tabular}{ l  l |l l}
    \toprule 
	\textbf{Element} & \textbf{Dimension} &  \textbf{Element} & \textbf{Dimension} \\
	\midrule
	$\tilde{T}_H$ to $T_H$          &  $1 \times 1 \times 4 = 4$ 
	& $\tilde{T}_H$ to $T_{t}$          &  $2 \times 1 \times 4 = 8$\\
	$\tilde{T}_D$ to $T_D$           &  $1 \times 3 \times 4 = 12$
	& $\tilde{T}_H$ to $T_{D}$          &  $2 \times 1 \times 4 = 8$\\
	$\tilde{T}_{E}$ to $T_{E}$      &  $1 \times 2 \times 4 = 8$ 
	&$\tilde{T}_D$ to $T_{t}$           &  $2 \times 3 \times 4 = 24$\\
	$\tilde{T}_{t}$ to $T_{t}$        &  $1 \times 3 \times 4 = 12$ 
	&$\tilde{T}_D$ to $T_{E}$           &  $2 \times 2 \times 4 = 16$\\
	& 
	& 	$\tilde{T}_{t}$ to $T_{E}$           &  $2 \times 2 \times 4 = 16$ \\
	\midrule
	Total & 36 & & 72 \\
    \bottomrule
  \end{tabular}
  \label{tbl:element-wise-feature}
\end{table}
\begin{table}[t]
\centering
\caption{Table-based search features used in various instantiations of our element-wise table matching framework. Element-wise and cross-element features are summarized in Table~\ref{tbl:element-wise-feature}, table feature are listed in Table~\ref{tbl:features}.}
\label{tbl:crab}
\begin{tabular}{cccc}
    \toprule  
	\multirow{2}{*}{\textbf{Method}}& \multicolumn{3}{c}{\textbf{Table similarity features}} \\
	\cline{2-4}
	                     & \textbf{Element-wise} & \textbf{Cross-element}  & \textbf{Table features}  \\
	\midrule
     STR-t1 & $\surd$ &   & \\
     STR-t2 & $\surd$ &   & $\surd$ \\
     STR-t3 & & $\surd$ & $\surd$ \\ 
     STR-t4 & $\surd$ & $\surd$ & $\surd$ \\
    \bottomrule
  \end{tabular}
\end{table}
\noindent
We present four specific instantiations of our table matching framework, by considering various combinations of the three main groups of features.  These instantiations are labeled as \emph{STR-t1} .. \emph{STR-t4} and are summarized in Table~\ref{tbl:crab}.

\section{Test Collection}
\label{sec:testcoll}

We introduce our test collections, including the table corpus, test and development query sets, and the procedure used for obtaining relevance assessments.

\subsection{Table Corpus}
\label{sec:testcoll:tablecorpus}

We use the WikiTables corpus~\citep{Bhagavatula:2015:TEL}, which comprises 1.6M tables extracted from Wikipedia (dump date: 2015 October). 
The following information is provided for each table: table caption, column headings, table body, (Wikipedia) page title, section title, and table statistics like number of headings rows, columns, and data rows.
We further replace all links in the table body with entity identifiers from the DBpedia knowledge base (version 2015-10) as follows.  For each cell that contains a hyperlink, we check if it points to an entity that is present in DBpedia.  If yes, we use the DBpedia identifier of the linked entity as the cell's content; otherwise, we replace the link with the anchor text, i.e., treat it as a string.

\subsection{Keyword-based Search}

\subsubsection{Queries}

We sample a total of 60 test queries from two independent sources (30 from each):
(1) \emph{Query subset 1 (QS-1)}: \citet{Cafarella:2009:DIR} collected 51 queries from Web users via crowdsourcing (using Amazon's Mechanical Turk platform, users were asked to suggest topics or supply URLs for a useful data table). 
(2) \emph{Query subset 2 (QS-2)}: \citet{Venetis:2011:RST} analyzed the query logs from Google Squared (a service in which users search for structured data) and constructed 100 queries, all of which are a combination of an instance class (e.g., ``laptops'') and a property (e.g., ``cpu'').
Following~\citep{Bhagavatula:2013:MEM}, we concatenate the class and property fields into a single query string (e.g., ``laptops cpu''). 
Table~\ref{tbl:queries} lists some examples. 

\begin{table}[t]
\centering
\caption{Example keyword-based search queries from our query set.}
\begin{tabular}{p{5.5cm}p{5.5cm}}
	\toprule
	\textbf{Queries from~\citep{Cafarella:2009:DIR}}  & \textbf{Queries from~\citep{Venetis:2011:RST}} \\
	\midrule
	video games  & asian coutries currency \\
	us cities & laptops cpu\\
	kings of africa & food calories\\
	economy gdp & guitars manufacturer\\
	fifa world cup winners&clothes brand\\	
	\bottomrule
\end{tabular}
\label{tbl:queries}
\end{table}
\label{sec:data:query}

\subsubsection{Relevance Assessments}

We collect graded relevance assessments by employing three independent (trained) judges.  
For each query, we pool the top 20 results from five baseline methods (cf. Sect.~\ref{sec:eval:baselines}), using default parameter settings. (Then, we train the parameters of those methods with help of the obtained relevance labels.)
Each query-table pair is judged on a three point scale: 0 (non-relevant), 1 (somewhat relevant), and 2 (highly relevant).
Annotators were situated in a scenario where they need to create a table on the topic of the query, and wish to find relevant tables that can aid them in completing that task.
Specifically, they were given the following labeling guidelines:
(i) a table is \emph{non-relevant} if it is unclear what it is about (e.g., misses headings or caption) or is about a different topic; 
(ii) a table is \emph{relevant} if some cells or values could be used from this table; and 
(iii) a table is \emph{highly relevant} if large blocks or several values could be used from it when creating a new table on the query topic. 
We take the majority vote as the relevance label; if no majority agreement is achieved, % (i.e., all three scores are different), 
we take the average of the scores as the final label.
To measure inter-annotator agreement, we compute the Kappa test statistics on test annotations, which is 0.47. According to~\citep{Fleiss:1971:MNS}, this is considered as moderate agreement. 
For each input query, there are on average 7.9 relevant and 6.28 highly relevant results.

%In total, 3120 query-table pairs are annotated as test data. 
%Out of these, 377 are labeled as highly relevant, 474 as relevant, and 2269 as non-relevant.

\subsection{Table-based Search}

\subsubsection{Queries}
\begin{figure*}[t]
   \centering
   \includegraphics[width=0.55\textwidth]{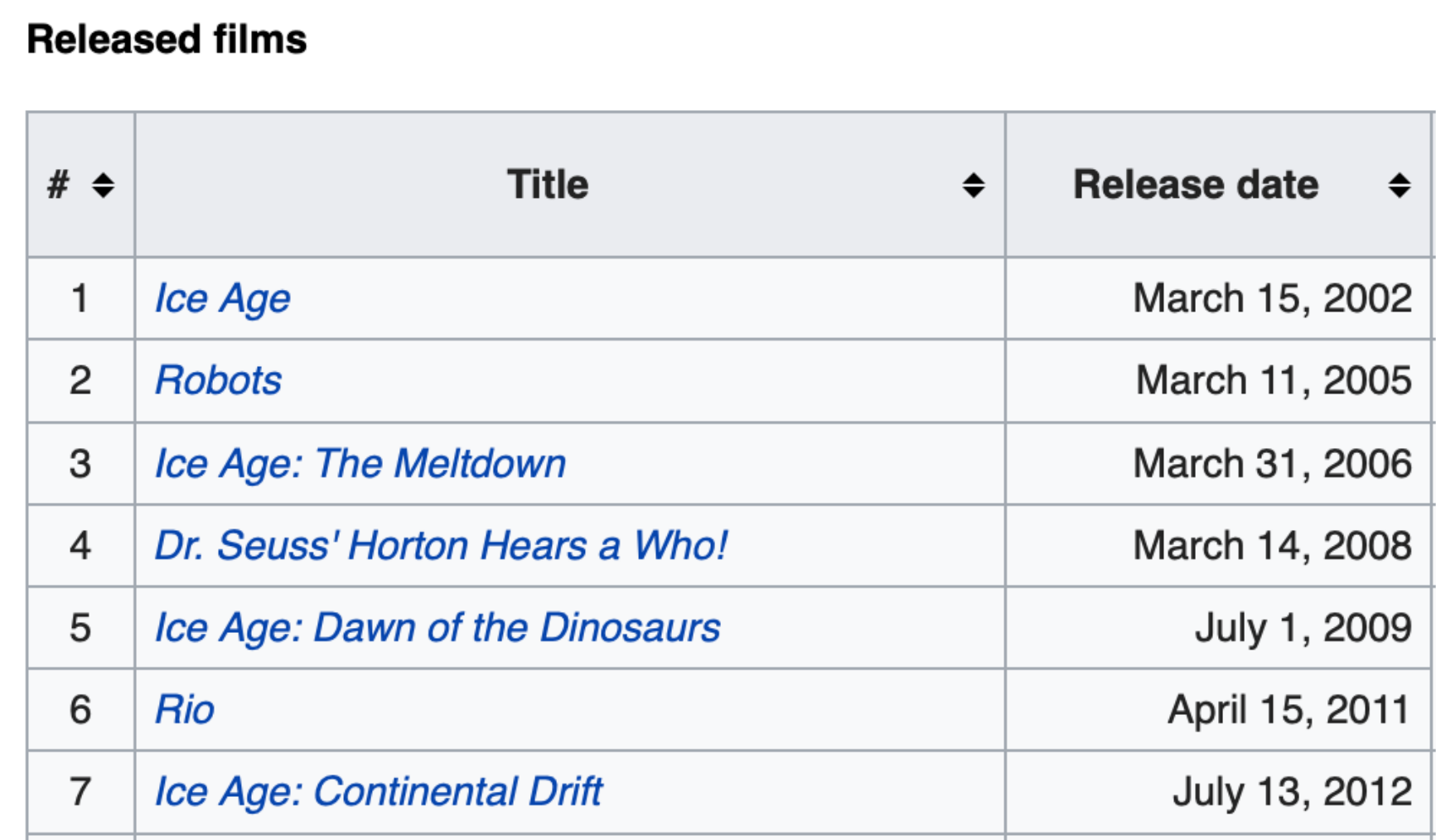} 
   \caption{An example query for table-based search from our query set.}
\label{fig:qbt_q}
\end{figure*}

We sample 50 Wikipedia tables from the table corpus to be used as queries.  Each table is required to have at least five rows and three columns~\citep{Zhang:2017:ESA}.  These tables cover a diverse set of topics, including sports, music, films, food, celebrities, geography, and politics.
See Figure~\ref{fig:qbt_q} as an example.

\subsubsection{Relevance Assessments}
\label{subsec:testcoll:rel}

Ground truth relevance labels are obtained as follows.  For each input table, three keyword queries are constructed: (i) caption, (ii) table entities (entities from table plus the entity corresponding to the Wikipedia page in which the table is embedded), and (iii) table headings.  Each keyword query is used to retrieve the top 150 results, resulting in at most 450 candidate tables for each query table.  
All methods that are compared in the experimental section operate by reranking these candidate sets.
For each method, the top 10 results are manually annotated.

Each query-table pair is judged on a three point scale: non-relevant (0), relevant (1), and highly relevant (2).  A table is highly relevant if it is about the same topic as the input table, but contains additional novel content that is not present in the input table.
A table is relevant if it is on-topic, but it contains limited novel content; i.e., the content largely overlaps with that of the input table.\footnote{We note that the novelty requirement is not something we consider in our modeling.  The purpose behind the use of this particular term was to have a simple working definition of relevance that discourages finding duplicate content.}
Otherwise, the table is not relevant; this also includes tables without substantial content.
Three colleagues were employed and trained as annotators.  We take the majority vote as the relevance label; if no majority vote is achieved, the mean score is used as the final label.  To measure inter-annotator agreement, we compute the Fleiss Kappa test statistics, which is 0.6703. According to \citep{Fleiss:1971:MNS}, this is considered as  substantial agreement.
For each input table, there are on average 7.28 relevant and 4.18 highly relevant results.

\section{Evaluation}
\label{sec:eval}

In this section, we list our research questions (Sect.~\ref{sec:eval:rq}), discuss our experimental setup (Sect.~\ref{sec:eval:setup}), introduce the baselines we compare against (Sect.~\ref{sec:eval:baselines}), and present our results (Sect.~\ref{sec:eval:results}).

\subsection{Research Questions}
\label{sec:eval:rq}

The research questions we seek to answer are as follows.

\begin{description}
	\item[RQ1] Can semantic matching improve retrieval performance?
	%How effective are the semantic representations?
	\item[RQ2] Which of the semantic representations is the most effective?
	\item[RQ3] Which of the similarity measures performs better?
	\item[RQ4] How much do different table elements contribute to retrieval performance?
%	\item[RQ4] Which of the two element-level matching strategies performs better, element-wise or cross-element? 
%	\item[RQ5] How much do different table elements contribute to retrieval performance for query-by-table?
\end{description}

\subsection{Experimental Setup} %Features and Supervised Learning}
\label{sec:eval:setup}

We evaluate table retrieval performance in terms of Normalized Discounted Cumulative Gain (NDCG) at cut-off points 5, 10, 15, and 20.  
Our main evaluation metric for keyword-based search is NDCG@20, while for table-based search it is NDCG@10.\footnote{This choice is determined by the depth of the pools used when performing the assessments for each task.  At these cutoffs, we have complete judgments for all methods that are being compared.}
We also report feature importance based on Gini score, which is a measure of feature contribution.  Specifically, it computes each feature's importance as the sum over the number of splits that include the feature, proportionally to the number of samples it splits~\citep{MenzeKMHBPH:2009:ACR}. Formally,
\begin{equation}
	g(\tau) = 1 - p_1^2 - p_0^2 ~,
\end{equation}
where $p = \frac{n_k}{n}$ is the fraction of the $n_k$ samples from class $k=\{0,1\}$ out of the total of $n$ samples at node $\tau$.
To test significance, we use a two-tailed paired t-test and write $\dag$/$\ddag$ to denote significance at the 0.05 and 0.005 levels, respectively.

Our implementations are based on the Nordlys toolkit~\citep{Hasibi:2017:NTE}. 
We use an inverted index as the main underlying data structure. 
Each table is stored in an inverted index, with title, table caption, headings, entities, and table data fields.  Additionally, all table-related data is concatenated and stored in a single ``catchall'' field.
%E.g., to parse terms in attribute values, we remove stopwords and HTML markup, and lowercase tokens. 
All fields are parsed with stopword removal and stemming using Elasticsearch.

Many of our features involve external sources, which we explain below. 
To compute the entity-related features (i.e., features in Table~\ref{tbl:features} as well as the features based on the bag-of-entities representations in Table~\ref{tbl:features2}), we use entities from the DBpedia knowledge base that have an abstract (4.6M in total). 
The table's Wikipedia rank (yRank) is obtained using Wikipedia's MediaWiki API. 
The PMI feature is estimated based on the ACSDb corpus~\citep{Cafarella:2008:URW}.
For the distributed representations, we take pre-trained embedding vectors, as explained in Sect.~\ref{sec:sem:embeddings}.

\subsection{Implementation of Baselines}
\label{sec:eval:baselines}

\subsubsection{Baselines for Keyword-based Search}
We implement four baseline methods from the literature.

\begin{description}
	\item[Single-field document ranking] In \citep{Cafarella:2009:DIR, Cafarella:2008:WEP} tables are represented and ranked as ordinary documents. Specifically, we use Language Models with Dirichlet smoothing, and optimize the smoothing parameter using a parameter sweep.

	\item[Multi-field document ranking] \citet{Pimplikar:2012:ATQ} represent each table as a fielded document, using five fields: Wikipedia page title, table section title, table caption, table body, and table headings.  We use the Mixture of Language Models approach~\citep{Ogilvie:2003:MLM} for ranking.  Field weights are optimized using the coordinate ascent algorithm; smoothing parameters are trained for each field individually. 

	\item[WebTable] 
The method by \citet{Cafarella:2008:WEP} uses the features in Table~\ref{tbl:features} with \citep{Cafarella:2008:WEP} as source.  Following~\citep{Cafarella:2008:WEP}, we train a linear regression model with 5-fold cross-validation.

	\item[WikiTable] The approach by \citet{Bhagavatula:2013:MEM} uses the features in Table~\ref{tbl:features} with \citep{Bhagavatula:2013:MEM} as source.  We train a Lasso model with coordinate ascent with 5-fold cross-validation.

\end{description}
Additionally, we present a learning-to-rank approach using a rich set of features as a strong baseline:
\begin{description}
	\item[LTR-k] It uses the full set of features listed in Table~\ref{tbl:features}.  We employ pointwise regression using the Random Forest algorithm.\footnote{We also experimented with  Gradient Boosting regression and Support Vector Regression, and observed the same general patterns regarding feature importance. However, their overall performance was lower than that of Random Forests. We note that further improvements may be obtained by using other machine learning algorithms, e.g., LambdaMART, but this exploration is outside our scope.}
We set the number of trees to 1000 and the maximum number of features in each tree to 3.  We train the model using 5-fold cross-validation (w.r.t. NDCG@20). %; reported results are averaged over 5 runs.
\end{description}
The baseline results are presented in Table~\ref{tbl:results1}. 
It can be seen from this table that our LTR-k baseline (row five) outperforms all existing methods from the literature; the differences are substantial and statistically significant. 
Therefore, in the remainder of this paper, we shall compare against this strong baseline, using the same learning algorithm (Random Forests) and parameter settings.
We note that our emphasis is on the semantic matching features and not on the supervised learning algorithm.

\subsubsection{Baselines for Table-based Search}
We implement eight existing methods from literature as baselines. 
%The table elements used in these methods are listed in Table~\ref{tbl:baseline_method}.

\begin{description}
	\item[Keyword-based search using $T_E$] 
 The candidate table's score is computed by taking the terms from $\tilde{T}_E$ as the keyword query~\citep{Ahmadov:2015:THI}. This method queries an index of the table corpus against the table entities.
	\item[Keyword-based search using $T_H$]	\citet{Ahmadov:2015:THI} also use table headings as keyword queries. This method queries an index of the table corpus against the table headings.
	\item[Keyword-based search using $T_c$]	 Additionally, in this paper we also consider using the table caption as a query. This method searches against both the caption and catchall fields.
	\item[Mannheim Search Join Engine]	All candidate tables are scored against the input table using the FastJoin matcher~\citep{Wang:2011:FEM}. The edit distance threshold is set to $\delta=0.8$.
	\item[Schema complement]	 \citet{DasSarma:2012:FRT} aggregate the benefits of adding additional attributes from candidates tables to input tables as the matching score. The heading frequency statistics is calculated based on the Wikipedia table corpora and the heading similarity is aggregated using average.
	\item[Entity complement]	 The aggregated scores of the benefits of adding additional entities is taken as the matching score~\citep{DasSarma:2012:FRT}. WLM is based on entity out-links.  The data similarity threshold is set the same  as for string comparison, i.e., $\delta=0.8$. 
	\item[Nguyen et al.]	 Headings and table data are represented as term vectors for table matching in~\citep{Nguyen:2015:RSS}. The smoothing parameter value is taken from~\citep{Nguyen:2015:RSS} to be $\alpha=0.5$.
	\item[InfoGather] Element-wise similarity across four table elements: table data, column values, page title, and column headings are combined by training a linear regression scorer~\citep{Yakout:2012:IEA}.	InfoGather is trained using linear regression with coordinate ascent.	
\end{description}
Additionally, we present a learning-to-rank approach in two variants, using different sets of features, as strong baselines:

\begin{description}
	\item[LTR-t1] It uses all the table matching scores in Sect.~\ref{sec:fmt}. We take the same training mechanism and configuration as for LTR-k (that is, we optimize for NDCG@10).
	\item[LTR-t2] Compared to LTR-t1, it additionally considers table features, for both the query and candidate tables, and follows the same training step.
\end{description}

%\begin{table}[t]
%  \centering
%  \caption{Table elements used in existing methods.}
%  \begin{tabular}{lc@{~~~~~}c@{~~~~~}c@{~~~~~}c@{~~~~~}c}
%    \toprule
%    \textbf{Method}                                                 & \textbf{$T_c$}    & \textbf{$T_p$}  & \textbf{$T_E$} & \textbf{$T_H$} & \textbf{$T_D$} \\
%    \midrule
%    Keyword-based search using $T_E$~\citep{Ahmadov:2015:THI} &         &         & $\surd$ &         &         \\
%    Keyword-based search using $T_H$~\citep{Ahmadov:2015:THI} &         &         &         & $\surd$ &         \\
%    Keyword-based search using $T_c$ & $\surd$ &         &         &         &         \\
%    Mannheim Search Join Engine~\citep{Wang:2011:FEM}        &         &         &         & $\surd$ &         \\
%    Schema complement~\citep{DasSarma:2012:FRT}                     &         &         & $\surd$ & $\surd$ &         \\
%    Entity complement~\citep{DasSarma:2012:FRT}                    &         &         & $\surd$ &         &         \\
%    \citet{Nguyen:2015:RSS}                        &         &         &         & $\surd$ & $\surd$ \\
%    InfoGather~\citep{Yakout:2012:IEA}                   &         & $\surd$ &         & $\surd$ & $\surd$ \\
%    \bottomrule
%  \end{tabular}
%  \label{tbl:baseline_method}
%\end{table}
%
\noindent
%The experimental configurations of the various methods are as follows.
%
%For \emph{keyword-based search}, the $T_E$ and $T_H$ methods query an index of the table corpus against the respective fields, while the $T_c$ variant searches against both the caption and catchall fields.
%; all the three methods use BM25. 
%For the \emph{Mannheim Search Join Engine}, the edit distance threshold is set to $\delta=0.8$. For \emph{schema complement}, the heading frequency statistics is calculated based on the Wikipedia table corpora and the heading similarity is aggregated using average. For \emph{entity complement}, WLM is based on entity out-links.  The data similarity threshold is set the same  as for string comparison, i.e., $\delta=0.8$. To parse terms in attribute values, we remove stopwords and HTML markup, and lowercase tokens.  For \emph{Nguyen et al.}, the smoothing parameter value is taken from~\citep{Nguyen:2015:RSS} to be $\alpha=0.5$.  \emph{InfoGather} is trained using linear regression with coordinate ascent. Our \emph{LTR-t*} methods are trained using Random Forest Regression with 5-fold cross-validation; the number of trees is 1000 and the maximum number features is 3.
%
The first block in Table~\ref{tbl:results_crab} presents the evaluation results for the eight baselines for table-based search.
Among the three keyword-based search methods, which operate on a single table element (top 3 lines), the one that uses table headings as the keyword query performs the best, followed by table entities and table caption.
These unsupervised methods essentially treat partial tables (omitting rows and columns) as queries. % This a worthy trying to consider to extend them using STR in the future.
The methods in lines 4--8 consider multiple table elements; all of these outperform the best single-element method.  The approach that performs best among all, by a large margin, is InfoGather, which incorporates four different table elements.
Consequently, in our discussion below, we will focus exclusively on InfoGather as the best baseline from the literature.

\subsection{Experimental Results}
\label{sec:eval:results}

We now answer our research questions based on the results of our experiments.

\begin{table*}[t]
  \centering
  \caption{Keyword-based search performance. Significance is tested against LTR-k. Highest scores are in boldface.}
%  \footnotesize
  \begin{tabular}{ l  llll }
    \toprule 
	\textbf{Method} & \textbf{NDCG@5} & \textbf{NDCG@10} & \textbf{NDCG@15} & \textbf{NDCG@20} \\
	\midrule
	Single-field document ranking& 0.4315  & 0.4344  & 0.4586  & 0.5254  \\
	Multi-field document ranking & 0.4770 & 0.4860 & 0.5170 & 0.5473 \\
	WebTable~\citep{Cafarella:2008:WEP} & 0.2831  & 0.2992  & 0.3311  & 0.3726 \\
	WikiTable~\citep{Bhagavatula:2013:MEM} & 0.4903 & 0.4766 & 0.5062 & 0.5206 \\
	\midrule
	LTR-k  % (features from Table~\ref{tbl:features}) 
		& 0.5527 & 0.5456 & 0.5738 & 0.6031 \\
	\midrule
	STR-k  %Semantic table retrieval (features from Table~\ref{tbl:features} and Table~\ref{tbl:features2}) 
		& \textbf{0.5951} & \textbf{0.6293}$^\dag$ & \textbf{0.6590}$^\ddag$ & \textbf{0.6825}$^\dag$ \\
    \bottomrule
  \end{tabular}
  \label{tbl:results1}
\end{table*}

\begin{description}
	\item[RQ1] Can semantic matching improve retrieval performance?
\end{description}

%\paragraph{Keyword-based Table Search}
For keyword-based search, the last line of Table~\ref{tbl:results1} shows the results for our semantic table retrieval method (STR-k).  It combines the baseline set of features (Table~\ref{tbl:features}) with the set of novel semantic matching features (from Table~\ref{tbl:features2}, 12 in total).  
We find that these semantic features bring in substantial and statistically significant improvements over the strong LTR baseline.  %Thus, we answer \textbf{RQ1} positively. 
The relative improvements range from 7.6\% to 15.3\%, depending on the rank cut-off.

%\paragraph{table-based search}

\begin{table}[t]
  \centering
%  \small
  \caption{Table-based search performance. Significance is tested against InfoGather. Highest scores are in boldface.}
  \begin{tabular}{ l  ll }
    \toprule 
	\textbf{Method} & \textbf{NDCG@5} & \textbf{NDCG@10}  \\
	\midrule
		Keyword-based search using $T_E$  & 0.2001 & 0.1998 \\
	Keyword-based search using $T_H$  & 0.2318 & 0.2527 \\
	Keyword-based search using $T_c$  & 0.1369 & 0.1419 \\
	Mannheim Search Join Engine~\citep{Wang:2011:FEM}  & 0.3298 & 0.3131 \\
	Schema complement~\citep{DasSarma:2012:FRT}  & 0.3389 & 0.3418 \\
	Entity complement~\citep{DasSarma:2012:FRT}  & 0.2986 & 0.3093 \\
	\citet{Nguyen:2015:RSS}  & 0.2875 & 0.3007\\
	InfoGather~\citep{Yakout:2012:IEA}  & \textbf{0.4530} & \textbf{0.4686} \\
	\midrule	
	LTR-t1 (feats. from Sect.~\ref{sec:existing}) & 0.5382 & 0.5542 \\ 
	LTR-t2 (feats. from Sect.~\ref{sec:existing} and Table features in Tbl.~\ref{tbl:features}) & \textbf{0.5895$\dag$} & \textbf{0.6050$\dag$} 	\\
	\midrule
	STR-t1 & 0.5578 & 0.5672 \\
	STR-t2 & \textbf{0.6172$\ddag$} & \textbf{0.6267$\ddag$} \\
	STR-t3 & 0.5140 & 0.5282 \\	
	STR-t4 & 0.5804$\dag$ & 0.6027$\dag$ \\	
    \bottomrule
  \end{tabular}
  \label{tbl:results_crab}
\end{table}

%

\if 0
%
\begin{table}[t]
  \centering
%  \small
  \caption{Evaluation results for existing table-based methods from the literature. Best scores for each metric are boldfaced.}
  \begin{tabular}{ll@{~~}l@{~~}}
    \toprule 
	\textbf{Method} & \small{\textbf{NDCG@5}} & \small{\textbf{NDCG@10}} \\
	\midrule
	Keyword-based search using $T_E$  & 0.2001 & 0.1998 \\
	Keyword-based search using $T_H$  & 0.2318 & 0.2527 \\
	Keyword-based search using $T_c$  & 0.1369 & 0.1419 \\
	Mannheim Search Join Engine  & 0.3298 & 0.3131 \\
	Schema complement  & 0.3389 & 0.3418 \\
	Entity complement  & 0.2986 & 0.3093 \\
	Nguyen et al.  & 0.2875 & 0.3007\\
	InfoGather  & \textbf{0.4530} & \textbf{0.4686} \\	
%	\midrule
%	HCF-1 (feat. from Table~\ref{tbl:features_tsf}) & 0.5382 & 0.5542 \\ 
%	HCF-2 (feat. from Tables~\ref{tbl:features_tsf} and \ref{tbl:features_table}) & \textbf{0.5895$\dag$} & \textbf{0.6050$\dag$} \\
    \bottomrule
  \end{tabular}
  \label{tbl:results3}
\end{table}
%
\fi
%\begin{figure}[tbp]
%   \centering
%   \includegraphics[width=0.6\textwidth]{figures/CRAB2-VS-Info.png}   
%
%   \caption{Performance difference between InfoGather (baseline) and STR-t2 on the level of individual input tables. Positive bars indicate the advantage of STR-t2.}
%
%\label{fig:f_tlevel}
%\end{figure}
%
For table-based search, reported in Table~\ref{tbl:results_crab}, we first compare our strong baselines, LTR-*, against the best method from the literature, InfoGather.
LTR-t1, which combines all table similarity features from existing approaches, achieves 18.27\% improvement upon InfoGather in terms of NDCG@10, albeit the differences are not statistically significant.
LTR-t2 incorporates additional table features, which leads to substantial (29.11\% for NDCG@10) and significant improvements over InfoGather.
Next, we consider four specific instantiations of our table matching framework (cf. Table~\ref{tbl:crab}), which are presented in the bottom block of Table~\ref{tbl:results_crab}. 
Recall that STR-t1 employs only table similarity features, thus it is to be compared against LTR-t1.
STR-t2..4 additionally consider table features, which corresponds to the settings in LTR-t2.
We find that STR-t1 and STR-t2 outperform the respective LTR method, while STR-t4 is on par with it.  None of the differences between STR-t* and the respective LTR method are statistically significant.
%The best overall performer is LTR-t2, with a relative improvement of 36.2\% for NDCG@5 and 33.7\% for NDCG@10 over InfoGather.
%Figure~\ref{fig:f_tlevel} shows performance differences on the level of individual input tables between InfoGather and STR-t2.  Clearly, several tables are improved by a large margin, while only a handful of tables are affected negatively.
%
Taking a conservative stand, we might say that our semantic table matching methods are on par with the strong baselines (and not better than them).  This is already a great result for two reasons.  One is that the strong baselines rely on a large set of hand-crafted features and we can match that performance without extensive feature engineering.  The other is that the strong baselines already outperform the best reported approach in the literature by a substantial margin and deliver excellent performance (LTR-* vs. InfoGather).
%Overall, we can achieve the same performance as a state-of-the-art approach that relies on hand-crafted features (STR-t1 vs. LTR-t1 and STR-t2 vs. LRT-2).

%The summary of our findings thus far is that our semantic table element representation with element-wise matching is very effective.  We can   
%With that, we have accomplished our main research objective.

In summary, we answer RQ1 positively; semantic matching can improve retrieval performance substantially, both for keyword-based and for table-based search.

\begin{table*}[t]
  \centering
%\footnotesize
  \caption{Comparison of semantic features for keyword-based search, used in combination with baseline features for keyword-based search (from Table~\ref{tbl:features}), in terms of NDCG@20.}
%  Statistical significance is tested against the LTR baseline in Table~\ref{tbl:results1}.}
  \begin{tabular}{ l lllll }
    \toprule 
	\textbf{Sem. Repr.} 
	& \textbf{Early} 
	& \textbf{Late-max} 
	& \textbf{Late-sum} 
	& \textbf{Late-avg} 
	%& \textbf{Late} 
	& \textbf{ALL} \\
	\midrule
	Bag-of-entities % (B) 
		& 0.6754 %(+11.99\%) 
		& 0.6407 %(+6.23\%)$^\dag$ 
		& 0.6697 %(+11.04\%)$^\ddag$ 
		& 0.6733 %(+11.64\%)$^\ddag$ 
		%& 0.6638 (+10.06\%)$^\ddag$ 
		& 0.6696 \\%(+11.03\%)$^\ddag$ \\
%	Bag-of-categories %(C) 
%		& 0.6287 (+4.19\%) 
%		& 0.6245 (+3.55\%)  
%		& 0.6315 (+4.71\%)$^\dag$ 
%		& 0.6240 (+3.47\%)   
%		%& 0.6103 (+1.19\%)
%		& \cellcolor[gray]{0.85}0.6149 (+1.96\%) \\
	Word embeddings %(D) 
		& 0.6181 %(+2.49\%)  
		& 0.6328 %(+4.92\%)  
		& 0.6371 %(+5.64\%)$^\dag$ 
		& 0.6485 %(+7.53\%)$^\dag$ 
		%& 0.6451 (+6.96\%)$^\dag$  
		& 0.6588 \\%(+9.24\%)$^\dag$ \\
	Graph embeddings % (E) 
		& 0.6326 %(+4.89\%)  
		& 0.6142 %(+1.84\%)  
		& 0.6223 %(+3.18\%)
		& 0.6316 %(+4.73\%)  
%		%& 0.6357 (+5.41\%)  
		& 0.6340 \\% (+5.12\%) \\
	\midrule
	ALL
		& 0.6736 %(+11.69\%)$^\dag$ 
		& 0.6631 %(+9.95\%)$^\dag$ 
		& 0.6831 %(+13.26\%)$^\ddag$ 
		& 0.6809 %(+12.90\%)$^\ddag$  
		%& 0.6737 (+11.71\%)$^\ddag$ 
		& 0.6825  \\%(13.17\%)$^\ddag$ \\
	\bottomrule
  \end{tabular}
  \label{tbl:results2}
\end{table*}

\begin{description}
	\item[RQ2] Which of the semantic representations is the most effective?
\end{description}

For keyword-based search, Table~\ref{tbl:results2} reports on all combinations of semantic representations and similarity measures. 
%In the interest of space, we only report on NDCG@20; the same trends were observed for other NDCG cut-offs.  
%Cells with a white background show retrieval performance when extending the LTR baseline with a single feature.  
%Cells in the last row or column correspond to using a given semantic representation with different similarity measures (rows) or using a given similarity measure with different semantic representations (columns).  
%The first observation is that all features improve over the baseline, albeit not all of these improvements are statistically significant.
%
Concerning the comparison of different semantic representations, we find that bag-of-entities and word embeddings achieve significant improvements; see the rightmost column of Table~\ref{tbl:results2}.  
It is worth pointing out that for word embeddings the three similarity measures seem to complement each other, as their combined performance is better than that of any individual method.  It is not the case for bag-of-entities, where only one of the similarity measures (Late-max) is improved by the combination.
Overall, we find the bag-of-entities representation to be the most effective one.  The fact that this sparse representation outperforms word embeddings is regarded as a somewhat surprising finding, given that the latter has been trained on massive amounts of (external) data.

\if 0
\begin{table}[t]
  \centering
%  \small
  \caption{Comparison of semantic representations for table-based search. \todo{There seems to be some confusion with Tables 12 and 13. First, it is not clear why both of these tables are needed. Did the authors forget to remove Table 12? Second, the results in these tables don't match. In, particular, from Table 12 it follows that bag-of-entities representation works best for table-based search, despite what authors claim in line 47 of p. 21 and in the conclusion: Check.}}
  \begin{tabular}{ l  ll }
    \toprule 
	\textbf{Semantic Repr.} & \textbf{NDCG@5} & \textbf{NDCG@10}  \\
	\midrule
	Word embeddings & 0.3779 & 0.3906 \\
	Graph embeddings & 0.3012 & 0.3376 \\
	Bag-of-entities & 0.4484 & 0.4884 \\
	\midrule
	Combined & 0.5578 & 0.5672 \\
    \bottomrule
  \end{tabular}
  \label{tbl:results_repr}
\end{table}
\fi

%Now that we have assessed the overall effectiveness of our approach, let us turn to answering a series of more specific research questions.
%
For table-based search, Table~\ref{tbl:re} displays the results for each of the three semantic representations.  Among those, word-based performs the best, followed by bag-of-entities and graph embeddings. It is different from the findings on keyword-based search. 
The differences between bag-of-entities and word embeddings are significant ($p<$ 0.01), but not between the other pairs of representations.
It is worth pointing out that any of the three representations alone would deliver better performance than the best existing method in the literature, InfoGather (cf. Table~\ref{tbl:results2}). 
When combing all three semantic representations (line 4, which is the same as STR-t1 in Table~\ref{tbl:results_crab}), we obtain substantial and significant improvements ($p<$0.01) over each individual representation.  This shows the complimentary nature of these semantic representations.

In summary, bag-of-entities representations are the most effective for keyword-based search, and word embeddings work best for table-based search. 
However, these representations complement each other, and thus the combination of them performs best.

\begin{table*}[t]
  \centering
%\tiny
  \caption{Comparison of semantic features for table-based search, used in combination with table features (middle block in Table~\ref{tbl:features}), in terms of NDCG@10.}
%  \vspace*{-0.75\baselineskip}
  \begin{tabular}{ l lllll }
    \toprule 
	\textbf{Sem. Repr.} 
	& \textbf{Early} 
	& \textbf{Late-max} 
	& \textbf{Late-sum} 
	& \textbf{Late-avg} 
	%& \textbf{Late} 
	& \textbf{ALL} \\
	\midrule
	Bag-of-entities % (B) 
		&  0.5487
		&  0.5764
		&  0.5420
		&  0.5293
		&  0.5603\\
	Word embeddings %(D) 
		&  0.6173 
		&  0.5314 
		&  0.5512
		&  0.5337
		&  0.6048\\
	Graph embeddings % (E) 
		&  0.5072
		&  0.5676
		&  0.5270
		&  0.5313
		&  0.5559 \\
	\midrule
	ALL 
		& 0.6157
		& 0.6031
		& 0.5523
		& 0.5631
		& 0.6267 \\
	\bottomrule
  \end{tabular}
  \label{tbl:re}
\end{table*}

\begin{table*}[t]
%\small
\caption{Element-wise similarities for various semantic representations for table-based search.  Rows and columns corresponds to elements of the input and candidate tables, respectively. The evaluation metric is NDCG@10.  The best scores for each block are in boldface.}
\label{tbl:results_ewce}
\small
\begin{tabular}{llll l llll l llll}
    \toprule 
  	\multicolumn{4}{c}{Word embeddings} & &  
  	\multicolumn{4}{c}{Graph embeddings} & &   	
  	\multicolumn{4}{c}{Bag-of-entities} \\
  	\cline{1-4} \cline{6-9} \cline{11-14}
	 & \textbf{$T_{t}$} & \textbf{$T_{H}$} & \textbf{$T_{D}$} & & 
	 & \textbf{$T_{t}$} & \textbf{$T_{E}$} & \textbf{$T_{D}$} & & 
	 & \textbf{$T_{t}$} & \textbf{$T_{E}$} & \textbf{$T_{D}$}  \\
  	\cline{2-4} \cline{7-9} \cline{12-14}
	% W
	\textbf{$\tilde{T}_{t}$} & \textbf{0.2814} & 0.0261  & 0.0436 & &
	% G
	\textbf{$\tilde{T}_{t}$} & \textbf{0.2765} & 0.0546 & 0.0430 & & 
	% E
	\textbf{$\tilde{T}_{t}$} & \textbf{0.4796} & 0.0808 & 0.0644  \\
	% W
	\textbf{$\tilde{T}_{H}$} & 0.0336 & \textbf{0.1694} & 0.0288 & &
	% G
	\textbf{$\tilde{T}_{E}$} & \textbf{0.0700}  & 0.0679 & 0.0501 & &
	% E
	\textbf{$\tilde{T}_{E}$} & 0.0705 & 0.0617 & \textbf{0.0725}  \\
	% W
	\textbf{$\tilde{T}_{D}$} & 0.0509 & 0.0183 & \textbf{0.1276} & &
	% G
	\textbf{$\tilde{T}_{D}$} & \textbf{0.1012} & 0.0423 & 0.0259 & &
	% E
	\textbf{$\tilde{T}_{D}$} & \textbf{0.1052} & 0.0812 & 0.0610  \\
    \bottomrule
  \end{tabular}

\end{table*}
\begin{description}
	\item[RQ3] Which of the similarity measures performs better?
\end{description}

For keyword-based search, it is difficult to name a clear winner when a single semantic representation is used. The relative differences between similarity measures are generally small (below 5\%).
When all three semantic representations are used (bottom row in Table~\ref{tbl:results2}), we find that Late-avg and Late-sum achieve the highest overall improvement. 
Importantly, when using all semantic representations, all four similarity measures improve significantly and substantially over the baseline.  
We further note that the combination of all similarity measures do not yield further improvements over Late-sum or Late-avg.
Thus, we identify the late fusion strategy with sum or avg aggregation (i.e., Late-sum or Late-avg) as the preferred similarity method.

For table-based search, Early achieves the best performance, followed by Late-max, Late-avg, and Late-sum (bottom row in Table~\ref{tbl:re}).

In answer to RQ3, late fusion performs better for keyword-based search, while early fusion outperforms late fusion for table-based search. However, the differences are often small.  We hypothesize that early fusion performs better when obtaining representations ``symmetrically.'' Recall that table-based search takes entities for representing both queries and tables (or, more precisely, input and candidate tables) from the same sources symmetrically, while keyword-based search retrieves entities asymmetrically (the queries only get them from the knowledge base, while tables can also get entities from their body). However, this hypothesis would need to be carefully examined in a future study. Overall, the different similarity measures seem to complement each other, and taking their combination works well across both tasks.
% HCF-1 VS Info : 0.2124, 0.1884
% HCF-2 VS Info: 0.02044, 0.01368
% CRAB-1 VS Info: 0.09711, 0.1094
% CRAB-2 VS Info: 0.007334, 0.008293
% CRAB-3 VS Info: 0.3974, 0.3748
% CRAB-4 VS Info: 0.02617, 0.02046

% CRAB-1 VS HCF-1: 0.6825,  0.762
% CRAB-2 VS HCF-1: 0.07923, 0.07176
% CRAB-3 VS HCF-1: 0.6186,  0.5538
% CRAB-4 VS HCF-1: 0.3035,  0.1963
% CRAB-1 VS HCF-2: 0.4983,  0.3721
% CRAB-2 VS HCF-2: 0.4824,  0.5426
% CRAB-3 VS HCF-2: 0.11,    0.06432
% CRAB-4 VS HCF-2: 0.8055,  0.9462

\begin{description}
	\item[RQ4] How much do different table elements contribute to retrieval performance?
\end{description}

%\noindent
Note that this research question is specific to table-based search.  
Based on the results in Table~\ref{tbl:results_crab}, we observe that cross-element matching is less effective than element-wise matching (STR-t3 vs. STR-t2).  We also find that considering all the cross-element similarities actually hurts performance (STR-t4 vs. STR-t2).  In order to get a better understanding of how the element-wise and cross-element matching strategies compare against each other, we break down retrieval performance for all table element pairs according to the different semantic representations in Table~\ref{tbl:results_ewce}.  That is, we rank tables by measuring similarity only between that pair of elements (4 table similarity features in total).
Here, diagonal cells correspond to element-wise matching and all other cells correspond to cross-element matching.  We observe that element-wise matching works best across the board.  This is in line with our earlier findings, i.e., STR-t2 vs. STR-t3 in Table~\ref{tbl:results_crab}.
However, for graph embeddings and bag-of-entities representations, there are several cases where cross-element matching yields higher scores than element-wise matching.  Notably, input table data ($\tilde{T}_D$) has much higher similarity against the topic of the candidate table ($T_t$) than against its data ($T_D$) element, for both graph embeddings and bag-of-entities representations. This shows that cross-element matching does have merit for certain table element pairs. We perform further analysis in Sect.~\ref{sec:eval:analysis_fa}.

%\noindent
To explore the importance of table elements, we turn to Table~\ref{tbl:results_ewce} once again.
We first compare the results for element-wise similarity (i.e., the diagonals) and find that among the four table elements, table topic ($\tilde{T}_{t}\leftrightarrow T_{t}$) contributes the most and table data ($\tilde{T}_{D}\leftrightarrow T_{D}$) contributes the least.
Second, our observations for cross-element matching are as follows. In terms of feature importance (measured in terms of Gini score), using word embeddings, table data ($\tilde{T}_D$) is the most important element for the input table, while for the candidate table it is table topic ($T_t$).  Interestingly, for graph embeddings and bag-of-entities representations it is exactly the other way around: the most important input table element is topic ($\tilde{T}_t$), while the most important candidate table element is data ($T_D$).

\section{Analysis}
\label{sec:analysis}

We continue with further analysis of our results. 

\subsection{Further Analysis for Keyword-based Search}

\subsubsection{Features}
\label{sec:fa}
\begin{figure*}[!t]
   \centering
   \includegraphics[width=0.8\textwidth]{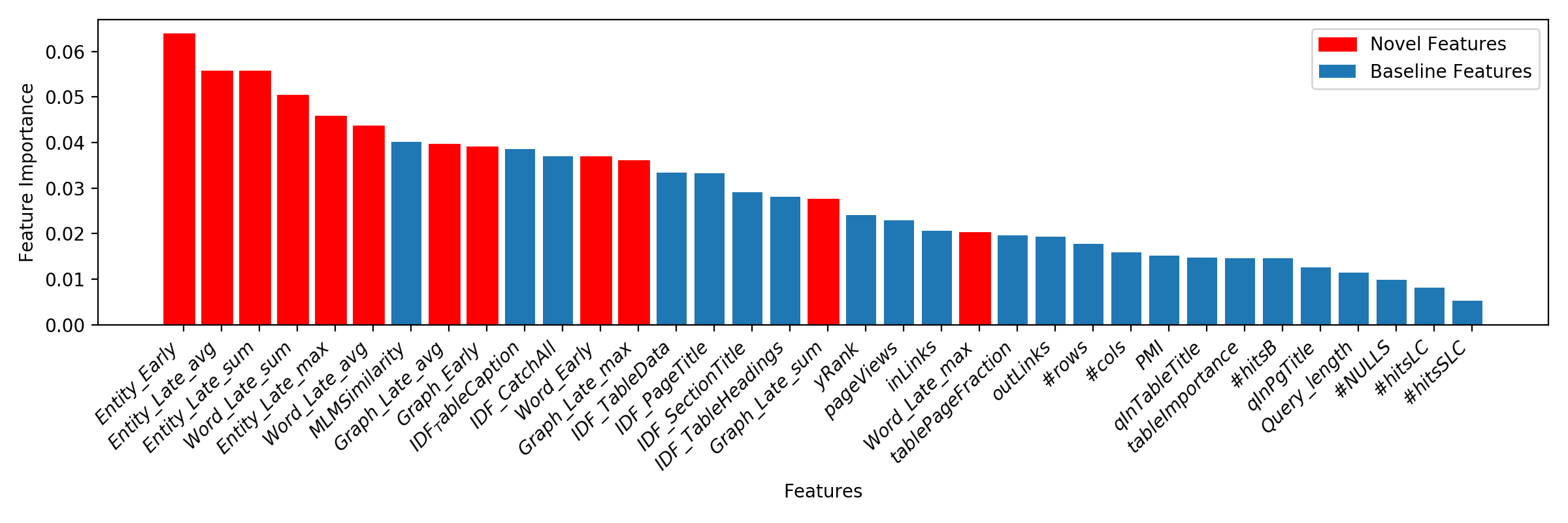} 
%   \vspace*{-\baselineskip}
   \caption{Normalized feature importance (measured in terms of Gini score) of keyword-based search.}
%   \vspace*{-0.5\baselineskip}
\label{fig:f_imp1}
\end{figure*}

Figure~\ref{fig:f_imp1} shows the importance of individual features for the keyword-based search task, measured in terms of Gini importance. The novel features are distinguished by color.
We observe that 8 out of the top 10 features are semantic features introduced in this paper.  
Additionally, we conduct a feature analysis based on retrieval performance. We find that most of the individual features help to improve table ranking, of which ``Graph\_Late\_sum'' improves the most in terms of NDCG@20 (0.11\%) and ``Word\_Late\_avg'' in terms NDCG@5 (3.8\%)''.

\begin{figure*}[t]
	\centering
	\begin{tabular}{ccc}
		\subfigure[Bag-of-entities]{\includegraphics[width=0.25\textwidth]{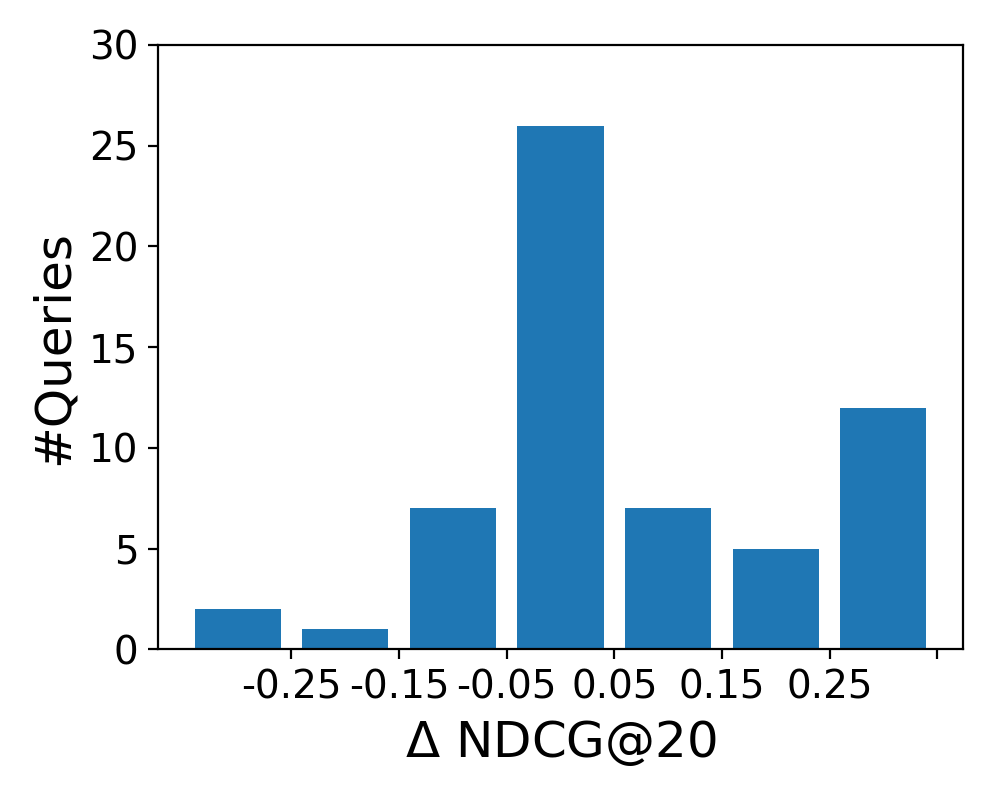}}
		&
%		\subfigure[Bag-of-categories]{\includegraphics[width=0.23\textwidth]{figures/queries_category.png}}
%		&
		\subfigure[Word embeddings]{\includegraphics[width=0.25\textwidth]{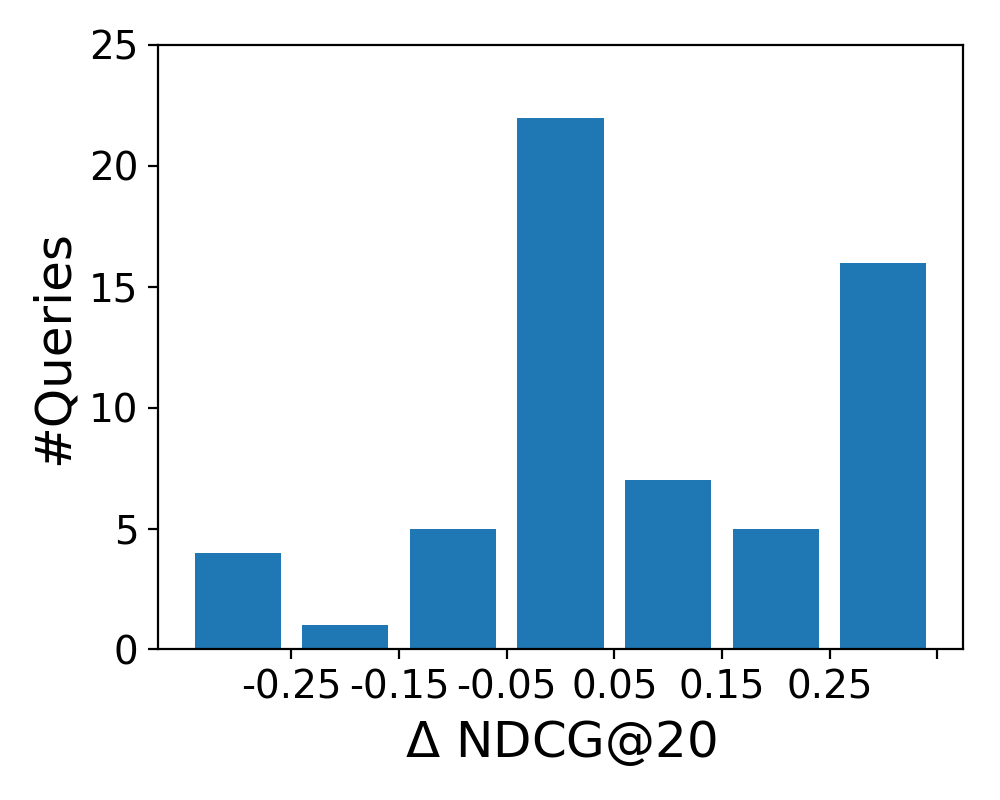}}
		&
		\subfigure[Graph embeddings]{\includegraphics[width=0.25\textwidth]{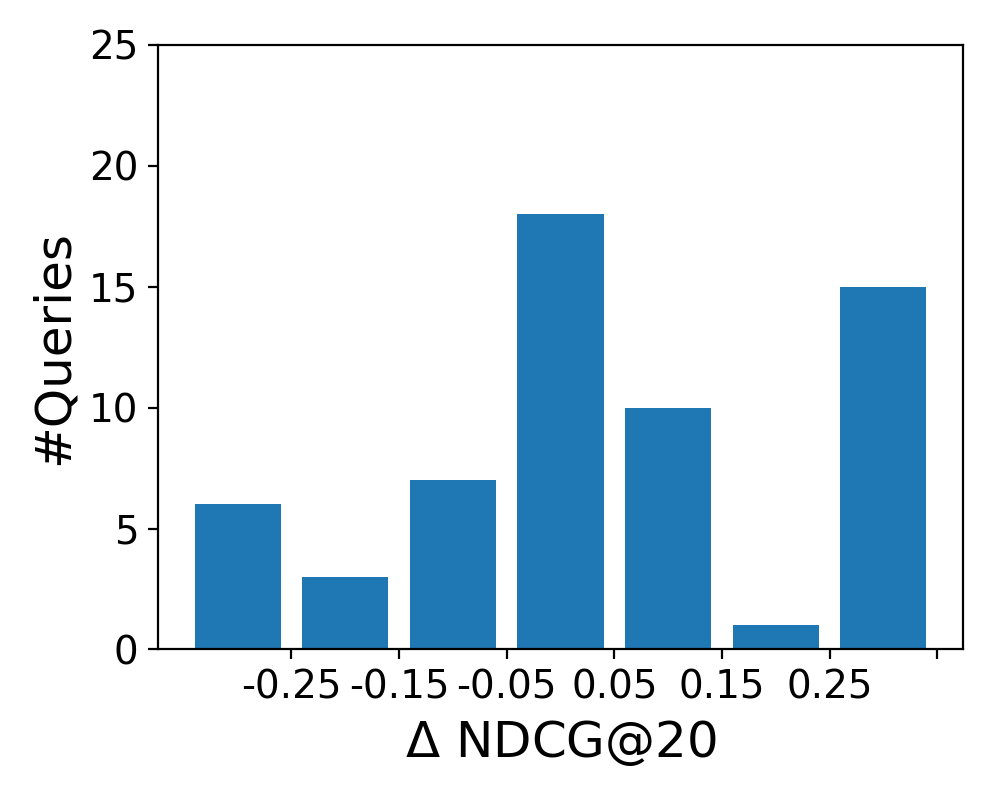}}
	\end{tabular}
%	\vspace*{-\baselineskip}
	\caption{Distribution of query-level differences between the LTR-k baseline and a given semantic representation for keyword-based search.}
	 \label{fig:ndcg_diff2}
\end{figure*}

\subsubsection{Semantic Representations}

To analyze how the three semantic representations affect retrieval performance on the level of individual queries, we plot the difference between the LTR-k baseline and each semantic representation in Figure~\ref{fig:ndcg_diff2}.  The histograms show the distribution of queries according to NDCG@20 score difference ($\Delta$): the middle bar represents no change ($\Delta<$0.05), while the leftmost and rightmost bars represents the number of queries that were hurt and helped substantially, respectively ($\Delta>$0.25).
We observe similar patterns for the bag-of-entities and word embeddings representations; the former has less queries that were significantly helped or hurt, while the overall improvement (over all topics) is larger.
We further note the similarity of the shapes of the distributions for graph embeddings. 

Looking at specific queries, we find one query that falls in the leftmost (i.e., largest performance drop) bucket in all the plots in Figure~\ref{fig:ndcg_diff2}: ``cereals nutritional value.''  This query only has a single highly relevant table according to the ground truth.
Further, there are five queries on which LTR consistently outperforms all semantic representations: ``Olympus digital SLRs,'' ``fuel consumption,'' ``Ibanez guitars,'' ``science discoveries,'' and ``lakes altitude''. What is common in these queries is that they are all very precisely worded.

%Overall, we observe that the majority of queries are improved and only few queries are affected negatively.
\subsubsection{Query Subsets}
\begin{figure}[t]
   \centering
   \includegraphics[width=0.5\textwidth]{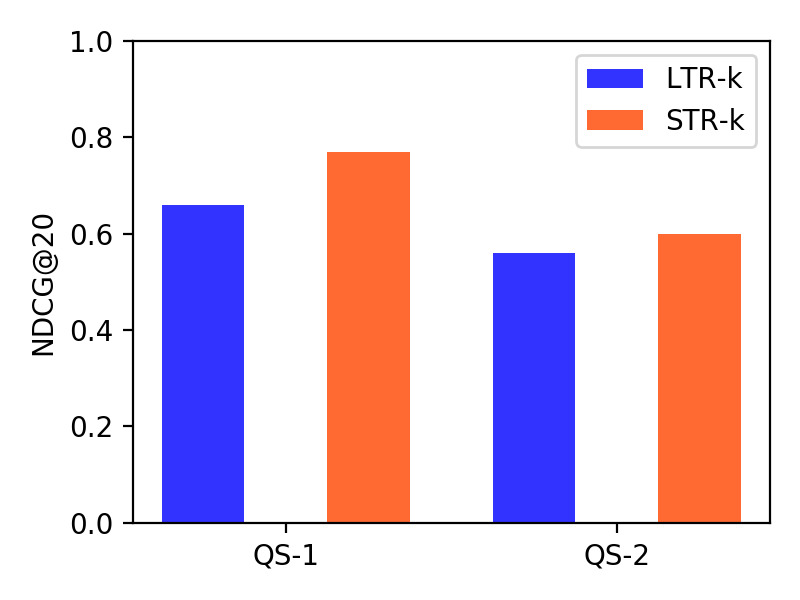}    
   \caption{Keyword-based search results, LTR-k baseline vs. STR-k, on the two query subsets in terms of NDCG@20.}
\label{fig:q1_VS_q2}
\end{figure}

On Figure~\ref{fig:q1_VS_q2}, we plot the results for the LTR-k baseline and for our STR-k method according to the two query subsets, QS-1 and QS-2, in terms of NDCG@20.  Generally, both methods perform better on QS-1 than on QS-2. This is mainly because QS-2 queries are more focused (each targeting a specific type of instance, with a required property), and thus are considered more difficult. 
Also, QS-1 has more relevant tables on average. Specifically, QS-1 has 8.4 relevant tables and 8.6 highly relevant tables, while QS-2 has 7.3 and 3.9, respectively.
Importantly, STR-k achieves consistent improvements over LTR-k on both query subsets.

\subsubsection{Individual Queries}
\begin{figure*}[t]
	\centering
	\begin{tabular}{cc}
		\subfigure[QS-1]{\includegraphics[width=0.5\textwidth]{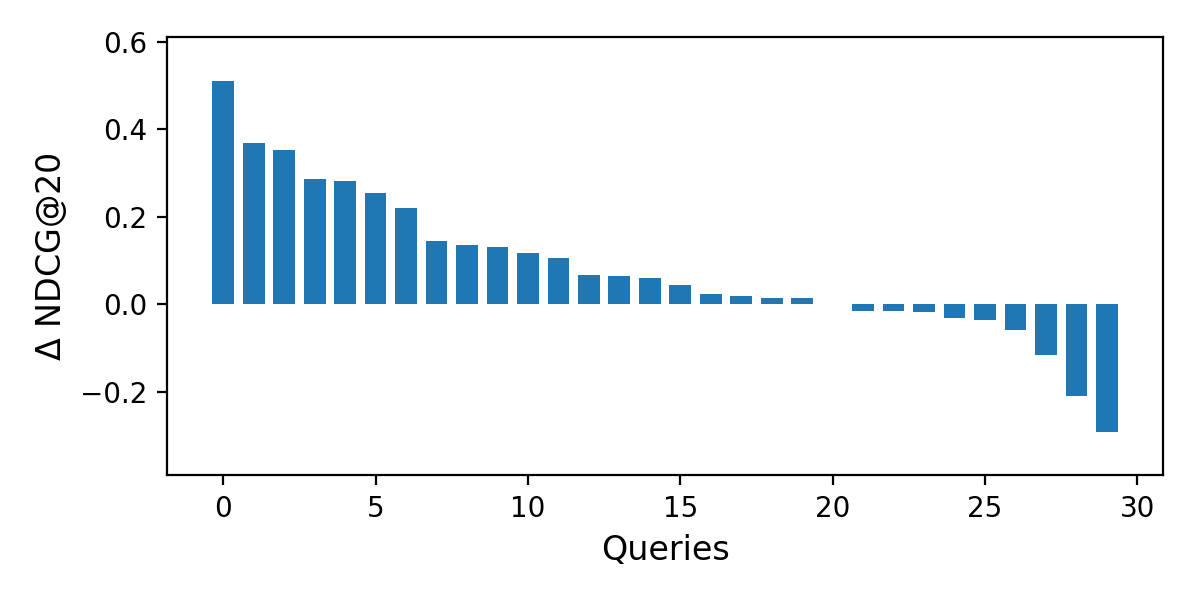}\label{fig:ndcg_diff_q1}}
		&
		\subfigure[QS-2]{\includegraphics[width=0.5\textwidth]{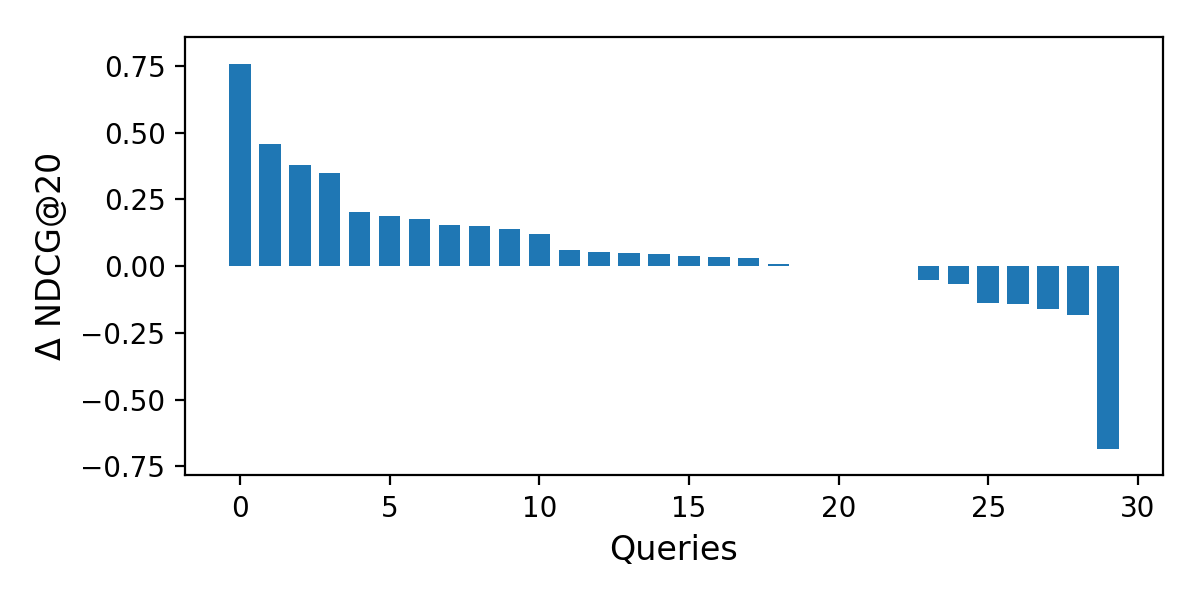}\label{fig:ndcg_diff_q2}}
	\end{tabular}
	\caption{Query-level differences on the two query subsets between the LTR-k baseline and STR-k. Positive values indicate improvements made by the latter.}
	 \label{fig:ndcg_diff_qs}
\end{figure*}

We plot the difference between the LTR-k baseline and STR-k for the two query subsets in Figure~\ref{fig:ndcg_diff_qs}.  Table~\ref{tbl:individual_queries} lists the queries that we discuss below.
The leftmost bar in Figure~\ref{fig:ndcg_diff_q1} corresponds to the query ``\emph{stocks}.'' For this broad query, there are two relevant and one highly relevant tables. LTR-k does not retrieve any highly relevant tables in the top 20, while STR-k manages to return one highly relevant table in the top 10.
The rightmost bar in Figure~\ref{fig:ndcg_diff_q1} corresponds to the query ``\emph{ibanez guitars}.'' For this query, there are two relevant and one highly relevant tables.  LTR-k produces an almost perfect ranking for this query, by returning the highly relevant table at the top rank, and the two relevant tables at ranks 2 and 4.  STR-k returns a non-relevant table at the top rank, thereby pushing the relevant results down in the ranking by a single position, resulting in a decrease of 0.29 in NDCG@20.

%\if 0
%
\begin{table}[t]
\centering
\caption{Example keyword-based search queries from our query set. Rel denotes table relevance level. LTR-k and STR-k refer to the positions on which the table is returned by the respective method. }
%\vspace*{-0.75\baselineskip}
\footnotesize
\begin{tabular}{lc@{~}c@{~}c@{~}}
	\toprule
	\textbf{Query} & Rel & LTR-k & STR-k  \\
	\midrule
	QS-1-24: \emph{stocks} \\
	~~~ Stocks for the Long Run / Key Data Findings: annual real returns & 2 & - & 6  \\
	~~~ TOPIX / TOPIX New Index Series & 1 & 9 & - \\
	~~~ Hang Seng Index / Selection criteria for the HSI constituent stocks & 1 & - & -  \\
	\midrule
	QS-1-21: \emph{ibanez guitars} \\ 
	 ~~~ Ibanez / Serial numbers & 2 & 1 & 2\\
	 ~~~ Corey Taylor / Equipment & 1 & 2 & 3 \\
	 ~~~ Fingerboard / Examples & 1 & 4 & 5 \\
	\midrule
	QS-2-27: \emph{board games number of players} \\
	~~~ List of Japanese board games & 1 & 13 & 1 \\ 
	 ~~~ List of licensed Risk game boards / Risk Legacy & 1 & - & 3 \\
	 
	 \midrule
	QS-2-21: \emph{cereals nutritional value} \\
	 ~~~ Sesame / Sesame seed kernels, toasted & 2 & 1 & 8 \\
	\midrule
	QS-2-20: \emph{irish counties area} \\
	~~~ Counties of Ireland / List of counties & 2 & 2 & 1 \\
	 ~~~ List of Irish counties by area / See also & 2 & 1 & 2 \\
	 ~~~ List of flags of Ireland / Counties of Ireland Flags & 2 & - & 3 \\
	 
	 ~~~ Provinces of Ireland / Demographics and politics & 1 & 4 & 4 \\
	~~~ Toponymical list of counties of the United Kingdom / Northern \dots & 1 & - & 7 \\
	~~~ Múscraige / Notes & 1 & - & 6 \\
	\bottomrule
\end{tabular}
\label{tbl:individual_queries}
\end{table}
%
%\fi
%
The leftmost bar in Figure~\ref{fig:ndcg_diff_q2} corresponds to the query ``\emph{board games number of players}.'' For this query, there are only two relevant tables according to the ground truth. STR-k managed to place them in the 1st and 3rd rank positions, while LTR-k returned only one of them at position 13th.
The rightmost bar in Figure~\ref{fig:ndcg_diff_q2} is the query ``\emph{cereals nutritional value}.'' Here, there is only one highly relevant result. LTR-k managed to place it in rank one, while it is ranked eighth by STR-k.
Another interesting query is ``\emph{irish counties area}'' (third bar from the left in Figure~\ref{fig:ndcg_diff_q2}), with three highly relevant and three relevant results according to the ground truth.  LTR-k returned two highly relevant and one relevant results at ranks 1, 2, and 4. STR-k, on the other hand, placed the three highly relevant results in the top 3 positions and also returned the three relevant tables at positions 4, 6, and 7.

Observing the individual queries, we find that, thanks to the semantic representations that can help bridge the vocabulary gap, STR-k is generally able to identify more relevant results than LTR-k. It should also be noted that for queries with only a handful relevant results, the rank position of a single table can make a large difference in NDCG.

\subsection{Further Analysis for Table-based Search}
\label{sec:eval:analysis}

Next, we perform further performance analysis on individual features and on input table size for table-based search.

\subsubsection{Feature Analysis}
\label{sec:eval:analysis_fa}

\begin{figure}[tbp]
   \centering
   \includegraphics[width=0.6\textwidth]{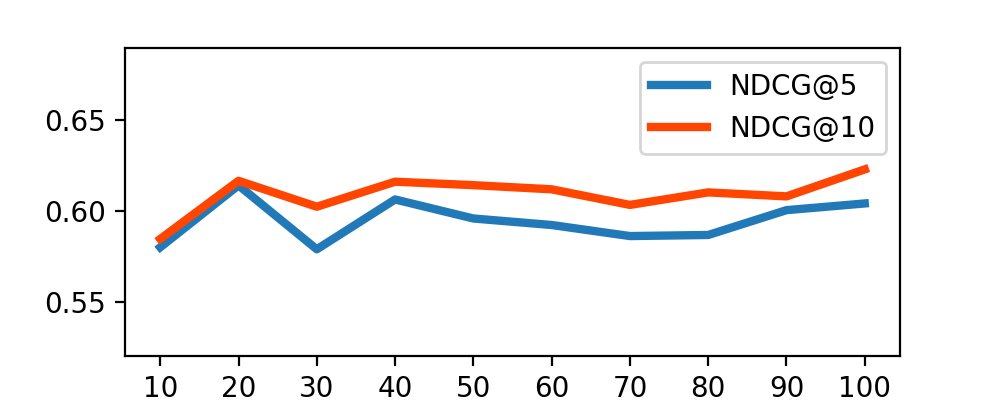} 

   \caption{Table-based search performance in terms of NDCG by incrementing the number of features used (features are ordered by Gini importance).}

\label{fig:top_k_feature}
\end{figure}
\begin{figure}[tbp]
   \centering
   \includegraphics[width=0.7\textwidth]{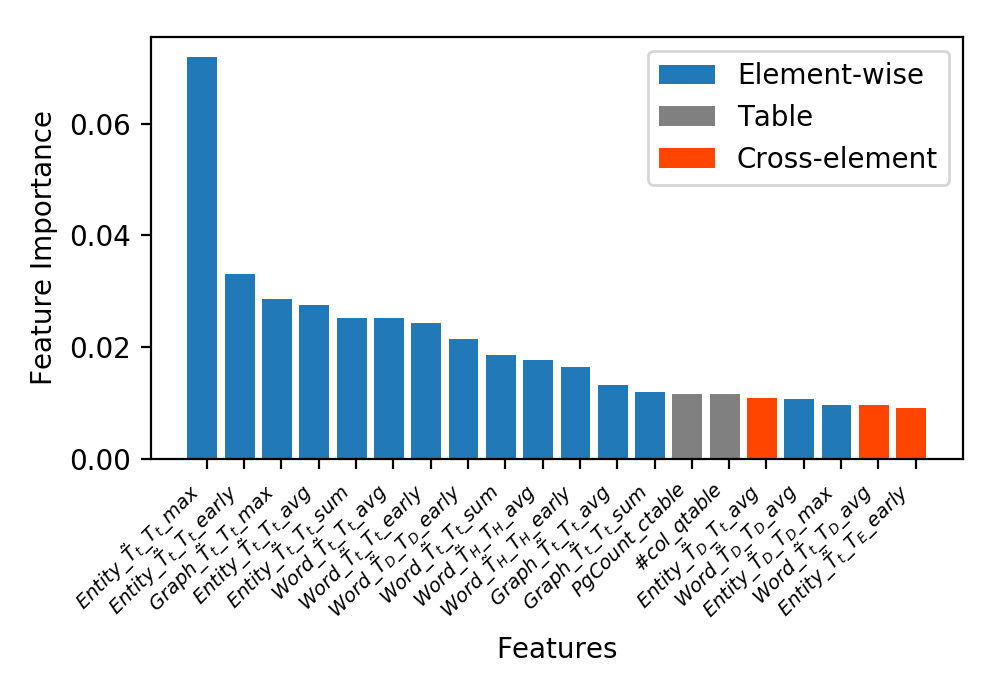}   

   \caption{Top-20 features in terms of Gini importance for table-based search.} %\todo{An interesting observation can be made from Figure 12, which shows that the most important features are w.r.t the table's topic.: check (what he says is hypothetical, no evidence)}}

\label{fig:f_imp}
\end{figure}
To understand the contributions of individual features, we first rank all features based on Gini importance. 
Then, we incrementally add features in batches of 10, and plot the corresponding retrieval performance in Figure~\ref{fig:top_k_feature}.
We observe that we can reach peak performance with using only the top-20 features.

Let us take a closer look at those top-20 features in Figure~\ref{fig:f_imp}.
We use color coding to help distinguish between the three main types of features: element-wise, cross-element, and table features.  Then, based on these feature importance scores, we revisit our research questions.
Concerning semantic representations, there are 8 word embedding, 7 entity embedding, and 3 graph embedding features in the top 20.  Even though there are slightly more features using word embedding than entity embeddings, the latter features are much higher ranked (cf. Fig.~\ref{fig:f_imp}).  Thus, the bag-of-entities semantic representation is the most effective one.
Comparing matching strategies, the numbers of element-wise and cross-wise features are 15 and 3, respectively.  This indicates a substantial advantage of element-wise strategies.  Nevertheless, it shows that incorporating the similarity between elements of different types can also be beneficial.  Additionally, there are 2 table features in the top 20.  
As for the importance of table elements, table topic ($T_{t}$) is clearly the most important one; 8 out of the top 10 features consider that element.
In summary, our observations based on the top-20 features are consistent with our earlier findings.

% MOVED FROM SEC 6 FOR BETTER LAYOUT
%
\begin{figure}[tbp]
   \centering
   \includegraphics[width=0.6\textwidth]{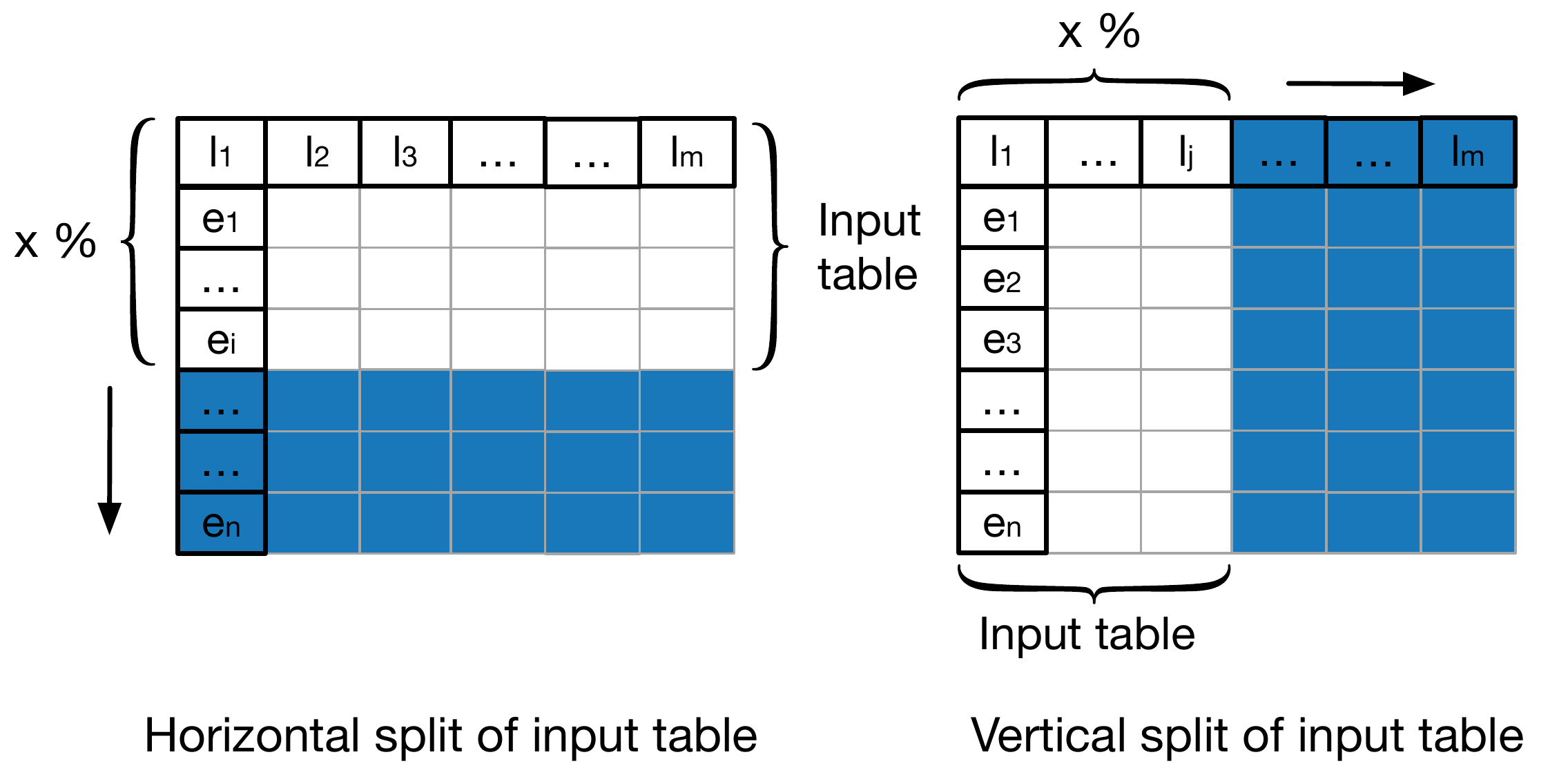}  

   \caption{Horizontal/vertically splitting of tables for performance analysis of table-based search.}

   \label{fig:query_split}
\end{figure}
%
% FIG FROM SECT 6
%
\begin{figure}[tbp]
   \centering
   \begin{tabular}{cc}
   \includegraphics[width=0.3\textwidth]{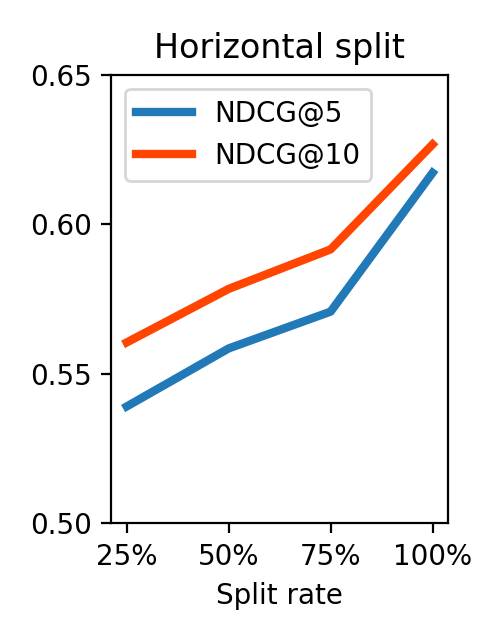} 
	&
   \includegraphics[width=0.3\textwidth]{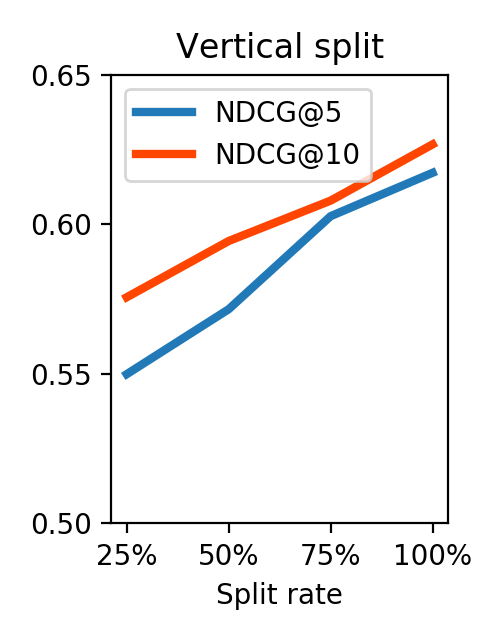} 	   	
   \end{tabular}

   \caption{Performance of STR-t2 with respect to (relative) input table size, by varying the number of rows (Left) or columns (Right).}

\label{fig:horizontal_split}
\end{figure}

\subsubsection{Input Table Size}

Next, we explore how the size of the input table affects retrieval performance. 
Specifically, we vary the input table size by splitting it horizontally (varying the number of rows) or vertically (varying the number of columns), and using only a portion of the table as input; see Fig.~\ref{fig:query_split} for an illustration.  We explore four settings by setting the split rate $x$ between 25\% and 100\% in steps of  25\%.
Figure~\ref{fig:horizontal_split} plots retrieval performance against input table size.
We observe that growing the table, either horizontally or vertically, results in proportional increase in retrieval performance.  This is not surprising, given that larger tables contain more information.  Nevertheless, being able to utilize this extra information effectively is an essential characteristic of our table matching framework.

\section{Conclusion}

In this paper, we have introduced and addressed the problem of \emph{table retrieval}: answering an information need with a ranked list of tables.
Specifically, we have studied this problem in two different flavors: \emph{keyword-based search}, where the information need is specified as a keyword query, and
\emph{table-based search}, where an existing table is used as input. 
The main contribution of this study is a \emph{semantic table retrieval} framework, which allows us to incorporate semantic matching into the task of table retrieval in a principled and effective way.  
In this framework, queries and tables can be represented using semantic concepts (bag-of-entities) as well as continuous dense vectors (word and graph embeddings) in a uniform way.  We have introduced multiple similarity measures for matching those semantic representations.
We have presented and experimentally compared a number of specific instantiations of the matching framework, depending on the type of the input query (keywords or table).
For evaluation, we have developed two purpose-built test collections based on Wikipedia tables.  We have considered a number of approaches from the literature for baseline comparison, and have also developed strong baselines for each task by combining elements from prior studies in feature-based supervised learning approaches.  These strong baselines represent substantial and significant improvements over all previous methods.  We have demonstrated that our semantic table retrieval approaches can either match (for table-based search) or significantly outperform (for keyword-based search) these strong baselines, while, unlike those, do not require extensive feature-engineering. 

There is a number of possible avenues to be considered in future work.  In this paper, we have resorted ourselves to simple pre-trained embeddings.  We conjecture that table retrieval would also benefit from advances in representation learning and neural language modeling, and possibly from task-specific fine-tuning.  Naturally, there are also other ways for aggregating word/entity embeddings into table element level embeddings, beyond what we explored in this paper.
Another direction concerns the choice of the table corpus.  In this work, we have worked with Wikipedia tables, which helped us to focus on modeling challenges, without being hindered by data quality issues.  It remains to be seen whether our findings generalize over to a more heterogeneous table collection with varying quality and imperfect entity annotations.
 
%We have found that entity-based representations are the most effective for keyword-based search, while word embeddings have worked best for table-based search.

%These similarity measures have been found to complement each other, and their combination has performed best.

%Another line of future work concerns the summarization of table search results, which has not been explored to date.

\bibliographystyle{ACM-Reference-Format}
\bibliography{00paper}

\end{document}